\documentclass[11pt,preprint]{aastex}
\usepackage{natbib}
\usepackage{longtable,lscape}
\usepackage{amssymb}
\usepackage{amsmath}
\usepackage{amsfonts}
\usepackage{xspace}

\newcommand{\HI}{\mbox{\sc H{i}}}
\newcommand{\HII}{\mbox{\sc H{ii}}}
\newcommand{\Fe}{\mbox [Fe{\sc{ii}}}
\newcommand{\Ne}{\mbox Ne{\sc{ii}}]}
\newcommand{\Si}{\mbox [Si{\sc{ii}}}
\newcommand{\Sulphur}{\mbox S{\sc{iii}}]}
\newcommand{\pahfit}{\mbox{\sc Pahfit}}
\newcommand{\kms}{\mbox{km s$^{-1}$}}

\newcommand{\htwo}{\mbox{$\rm{H}_{2}$}}

\slugcomment{to be submitted to ApJ}

\shorttitle{Shock-excited H$_2$ in Virgo}
\shortauthors{Wong et al.}

\begin{document}


\title{The search for shock-excited H$_2$ in Virgo spirals experiencing ram pressure stripping}


\author{O.\ Ivy Wong\altaffilmark{1,2}, Jeffrey D.\ P.\ Kenney\altaffilmark{2}, Eric J.\ Murphy\altaffilmark{3} and George  Helou\altaffilmark{4} }


\affil{$^{1}$CSIRO Astronomy \& Space Science, P.O. Box 76, Epping, New 
South Wales 1710, Australia}
\affil{$^{2}$Astronomy Department, Yale University, P.O. Box 208101 New 
Haven, CT 06520-8101, U.S.A.}
\affil{$^{3}$Observatories of the Carnergie Institution for Science, 813 Santa Barbara Street, Pasadena, CA 91101, U.S.A.}
\affil{$^{4}$California Institute of Technology, MC 405-47, 1200 East California Boulevard, Pasadena, CA 91125, U.S.A.}


\begin{abstract}  
We investigate the presence of shock-excited H$_2$ in four Virgo cluster galaxies that show clear 
evidence of ongoing ram pressure stripping.  Mid-infrared (MIR) spectral mapping of the rotational
 H$_2$  emission lines were performed using the Infrared Spectrograph (IRS) on board the 
{\it{Spitzer}} space telescope.  We target four regions along the leading side of galaxies 
where the intracluster medium (ICM) appears to be pushing back the individual galaxy's interstellar 
medium (ISM).  For comparison purposes, we also study two regions on the trailing side of these
 galaxies, a region within an edge-on disk and an extraplanar star-forming region.
We find a factor of 2.6 excess of warm \htwo/PAH  in our sample relative to the observed fractions in other 
nearby galaxies. We attribute the \htwo/PAH excess to contributions of shock-excited \htwo\ 
which is likely to have been triggered by ongoing ram pressure interaction in our sample galaxies.
Ram pressure driven shocks may also be responsible for the elevated ratios of \Fe/\Ne\  found in our
sample.  

\end{abstract}





\section{Introduction} 

%Shock-excited molecular hydrogen (\htwo) has been detected near low-luminosity AGN \citep[e.g.\ ][]{ogle07}
%and high-velocity tidal interactions between galaxies \citep[e.g.\ ][]{appleton06}.  \citet{roussel07} found that
%low-luminosity AGNs can be identified via an excess of \htwo\excitation.   
Ram pressure stripping is the removal of a galaxy's interstellar medium (ISM) by the
gaseous medium  through which it is moving. 
There is widespread evidence for ram pressure stripping in nearby clusters of galaxies, including
truncated disks of gas (neutral Hydrogen, HI), dust (far-infrared, FIR), and star formation (H$\alpha$)
within normal stellar disks \citep{koopmann04,chung09,cortese10},
and one-sided extraplanar tails of HI, H$\alpha$ and young stars \citep{kenney04,crowl05,chung07,cortese07,sun10,yagi10,smith10,merluzzi13}.
The loss of gas from ram pressure stripping is 
likely the main star formation quenching mechanism for spirals and dwarf galaxies in clusters,
and transforms galaxies from blue, late-type galaxies into red, early-type galaxies
\citep{koopmann04,crowl10,kormendy12,kenney14}.

In addition to removing gas from the galaxies, ram pressure may disturb the ISM gas which is not yet stripped.
The radio continuum emission provides evidence for this, as well as  evidence for ongoing ram pressure interactions.
Ridges of enhanced linearly polarized radio continuum emission are often observed 
along the leading sides of galaxies experiencing ram pressure \citep{vollmer04,vollmer07,vollmer08},
 presumably where the magnetized ISM is compressed.
Maps of the total radio continuum emission can show long tails on the trailing side \citep{gavazzi87,crowl05},
as well as local radio deficit regions on the leading side just beyond the 
ridges of enhanced radio polarization \citep{murphy09}.
Local radio deficit regions can be identified based on the good 
correlation between the FIR and the radio continuum emission in star-forming galaxies 
\citep{helou85,dejong85,yun01}.  The correlation is also good locally
except that the radio emission is more spread out than the FIR emission, since the relativistic electrons
can diffuse far from their acceleration sites \citep{hoernes98,murphy06,hughes06,murphy08,dumas11}.
While in undisturbed galaxies, the radio and smoothed FIR distributions are very similar,
\citet{murphy09} found that Virgo spirals experiencing ram pressure exhibit a local deficit of radio emission, with respect to
a smoothed FIR map, at the leading sides of the ram pressure interaction. 
This local radio deficit is due predominantly to a lack of relativistic electrons on the leading side,
probably because the magnetic fields are compressed and aligned along the leading edge,
preventing relativistic electrons from diffusing outwards in the direction of the ram pressure,
since they cannot cross the magnetic field lines \citep{pfrommer10}.

Interestingly, the galaxies which have the strongest local radio deficits have globally enhanced radio/FIR
ratios, suggesting that the radio emission is somehow enhanced by a ram pressure interaction \citep{murphy09}.
This global radio enhancement is not due simply to the radio tails, which are relatively weak in the Virgo galaxies,
but associated with the main body of the galaxy.
Evidence for enhanced radio emission in cluster galaxies has also been found by \citet{miller01} and \citet{reddy04}.
In one of the clearest cases of active ram pressure stripping (NGC 4522), 
\citet{vollmer04} measured flat radio spectral indices along the ridges of high polarization
which suggest the presence of cosmic ray electron acceleration \citep{murphy09}. Hence, the
excess radio emission could be caused by re-acceleration of cosmic ray electrons by shocks driven
throughout the ISM by ram pressure. To test the viability of this hypothesis, a measure of the
presence of shocks is needed in these galaxies with enhanced global radio emission.

We propose that cluster galaxies with excess non-thermal radio emission may contain warm
\htwo\ gas which has been shock-heated by the process of ram pressure stripping.   Shock-heated \htwo\ 
has been observed along a shock ridge resulting from a high-velocity collision between an
intruder  galaxy and the gas in the intragroup medium of Stephan's Quintet \citep{appleton06}.
 The shock-heated gas then cools and forms \htwo\ molecules which may be excited by low velocity, 
non-dissociative magnetohydrodynamic (MHD) shocks that dissipate most of the collision's 
mechanical energy.
Mid-infrared (MIR) observations of ESO 137-001  within the Abell 3627 cluster 
\citep{sivanandam09}  reveal a shock-excited warm molecular hydrogen tail corresponding to the X-ray and 
H$\alpha$ tails that have resulted from ram pressure stripping \citep{sun06,sun07}. Recently, ram-pressure
driven shocks were also observed in the A3558 cluster using optical line ratio diagnostics \citep{merluzzi13}.

%Low-luminosity AGNs can be identified via an excess of 
%\htwo\ excitation caused by shock heating \citep[e.g.\ ][]{roussel07}.  Although galaxies with AGNs have been observed 
%to have higher \htwo-to-aromatic band power ratios by up to a magnitude,  shock heated \htwo\ has  been found in 
%several galaxies not associated with AGNs or intergalactic gas clouds  \citep{appleton06,egami06}.  

%The temperature distribution estimations of the warm molecular gas phase ($T$$\sim$100--1000 K)
%can be made from observations of rotational transitions of molecular hydrogen since they account for a significant volume
%fraction of the molecular clouds \citep{roussel07}. %Table~\ref{h2props} lists the properties of the rotational transitions
%of \htwo\ which are observable in the mid-infrared (MIR) wavelengths.  

The purpose of this paper is to search for possible ram pressure-driven shock excitation of \htwo\ in 
four Virgo cluster spirals experiencing ongoing ram pressure stripping.  We perform MIR spectral 
mapping between the wavelengths of 5--38 $\mu$m using the Infrared Spectrograph \citep[IRS; ][]{houck04} 
on board the {\it{Spitzer}} Space Telescope. Section 2 describes our sample, observations and data 
processing.  The resulting spectral line measurements and rotational \htwo\ emission line maps are 
presented in Section 3.  %From the measured emission lines, we use MIR emission line ratio diagnostics to
%investigate  the ISM conditions of galaxies experiencing ram pressure stripping.

Section 4 examines the ISM conditions using the MIR emission in a variety of non-nuclear regions
sampled from four Virgo galaxies undergoing ram pressure stripping.  Section 4.1 investigates the
ratio of warm \htwo\ (rotational \htwo\ emission lines observed in the MIR) relative to the Polycyclic 
Aromatic Hydrocarbon (PAH) emission.  Since ongoing star formation and its associated \HII\  and photodissociation regions (PDRs) are 
responsible for a relatively uniform \htwo/PAH ratio, an increase in this ratio is indicative of \htwo\ 
heating via an additional mechanism such as shocks (possibly driven by the ram pressure
interaction).  Section 4.2 compares the MIR emission line ratios \Fe/\Ne\ and \Si/\Sulphur\ of our
sample to those from the SINGS survey.  Typically, observations of AGNs find enhanced ratios of \Fe/\Ne\ 
and \Si/\Sulphur\ relative to those of normal star-forming regions. Do galaxies experiencing ram pressure 
show similar levels of enhancement?  In Section 4.3,  we constrain the temperatures of the warm \htwo\ emission observed in 
our sampled regions using one-temperature and two-temperature dust models.  From this, we 
investigate whether the observed \htwo\ is warmer on the leading sides and cooler on the trailing
sides of galaxies falling into the ICM.  A summary of our results is detailed in Section 5.

%since an excess of non-thermal radio continuum emission was also found in most of the ram pressure stripped galaxies studied by \citet{murphy08}.  

% In addition to the molecular Hydrogen lines, 
%we will be able to use the ratios of other metal emission lines to estimate the hardness of the interstellar radiation field.  
%For example, the {$[$Ne \sc{iii}$]$/$[$Ne \sc{ii}$]$} ratio of star-forming regions show harder radiation fields for lower metal 
%abundance environments.  As such, {$[$Ne \sc{iii}$]$/$[$Ne \sc{ii}$]$} provide a good discriminant between 
%star-forming galaxies and galaxies with AGN.  

\section{Sample and observations}

%The purpose of our project is to study the shock-excited \htwo\ and the propagation of shocks in the ISM of a sample
%of galaxies experiencing ram-pressure stripping in the Virgo cluster.  
%We intend to observe the regions of galaxies known to be experiencing the greatest pressure (also known as the leading edge)
%as well as the trailing edges which are on the opposite side of the galaxy to the leading edge. We
%will describe our sample and observations in this section.

\subsection{Galaxy sample}
%We have selected four Virgo spirals (NGC 4330, NGC 4402, NGC 4501 and NGC 4522) with clear evidence for 
%strong ongoing ram pressure.  This evidence includes: (1) ridges of enhanced polarized radio continuum emission 
%\citep{vollmer04,vollmer07,vollmer08}; (2) one-sided gas and dust features \citep[e.g.\ ][]{chung07,wong10a};
%(3) extraplanar tails of \HI\ \citep{kenney04,crowl05,chung09,chung07}; and (4) young stellar populations 
%in their gas-stripped outer stellar disks \citep{crowl06,crowl08,abramson10}. 
%Figure~\ref{sdss} shows three-color optical images for our sample of galaxies from the Sloan Digital Sky 
%Survey (SDSS) overlaid with contours of neutral Hydrogen (\HI) gas from the VLA Imaging of Virgo in Atomic
%Hydrogen survey \citep[VIVA; ][]{chung09}.  The \HI\ gas morphology in each of these galaxies appears to be
%pushed from one side and lop-sided; while the optical stellar morphology appears completely unaffected.
 
We have selected four Virgo spirals (NGC 4330, NGC 4402, NGC 4501 and NGC 4522) with clear evidence for 
strong ongoing ram pressure.  Evidence that these galaxies have experienced ram pressure stripping
recently includes: (1) truncated gas disks within undisturbed stellar disks \citep{koopmann04,chung09}; 
(2) young stellar populations in their gas-stripped outer stellar disks \citep{crowl06,crowl08,abramson11};
 and (3) one-sided extraplanar tails of \HI\ and dust \citep{kenney04,crowl05,chung07,chung09,wong10b}.
These properties can persist after ram pressure diminishes, so partly reflect the stripping history of 
the galaxies.  We are most interested in measures of the {\it{current strength}} of ram pressure, since
this is likely to be related to the strength of any ram pressure-induced shocks.

Evidence of {\em{ongoing}} ram pressure is present in the form of: (1) one-sided ridges of enhanced
 polarized radio continuum \citep{vollmer04,vollmer07,vollmer08}; and (2) one-sided ``radio deficit regions''
 \citep{murphy09}.  The polarized ridges likely indicate ISM compression on the leading sides of
the ram pressure interactions \citep{vollmer04,vollmer07,vollmer08}.  Beyond the polarized ridges,
local radio deficit regions are found in ram pressure affected galaxies where the densities of 
cosmic ray electrons and magnetic fields are lower than in undisturbed galaxies.
%the radio emission is weaker than would be expected based on the observed FIR distribution and the
% FIR--radio correlation in galaxies \citep{murphy09}.  These ``local radio deficit regions'' are 
%likely to be regions where the densities of halo cosmic ray electrons and magnetic fields are lower
%than in an undisturbed galaxy, due to ram pressure.  
NGC 4402 and NGC 4522 each have strong polarized ridges and strong radio deficit regions.  NGC 4330
 has a strong radio deficit region but no prominent polarized ridge, perhaps because of an unfavorable 
viewing geometry \citep{vollmer11}.  NGC 4501 has a strong polarized ridge but is not in the same sample 
of \citet{murphy09} that measured the radio deficit regions.  Hence, all four of these galaxies 
have clear indications of strong ongoing ram pressure. 

\begin{figure}
\begin{center}
\includegraphics[scale=1.]{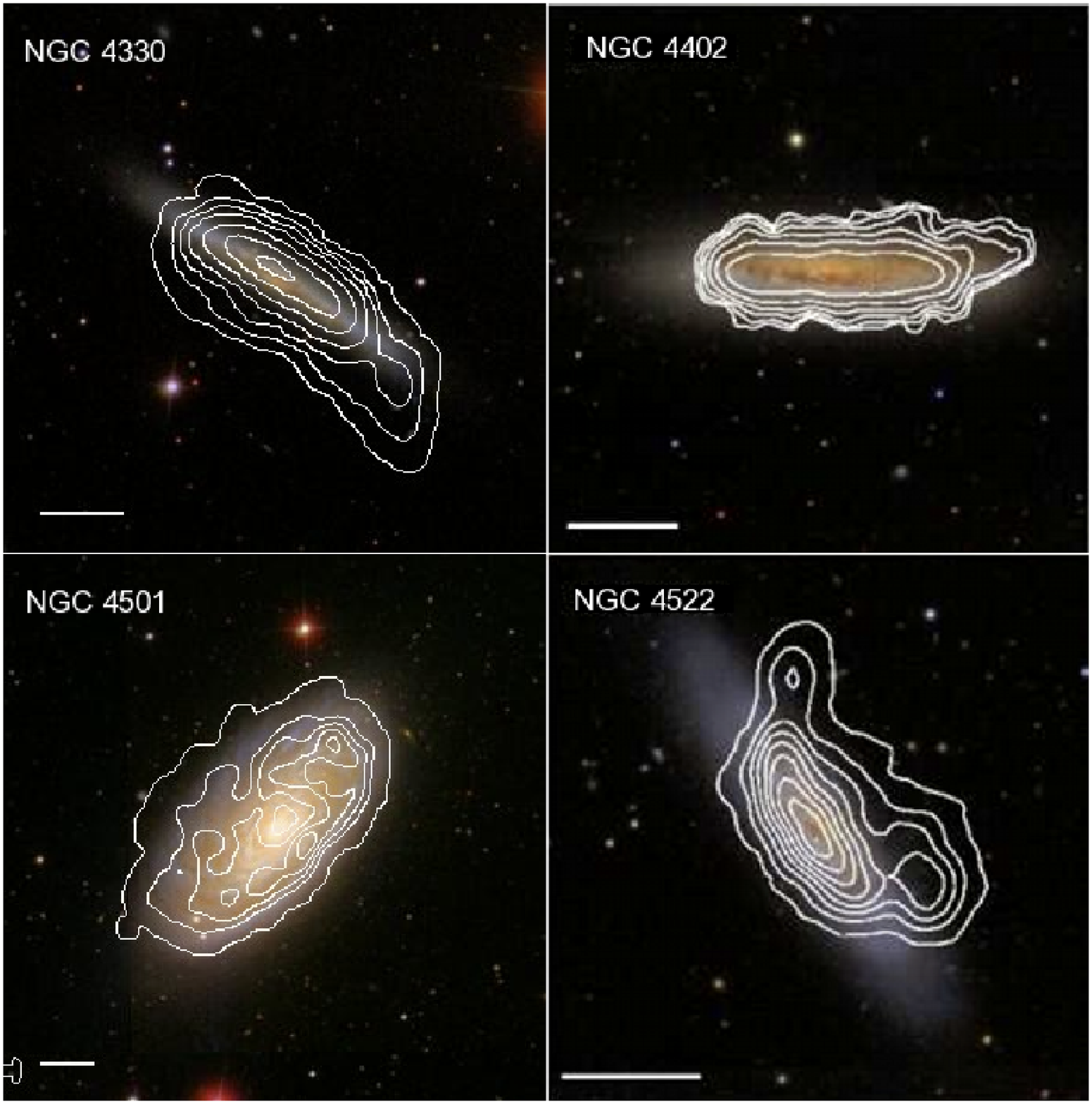}
\end{center}
\caption{Ram pressure stripping of the \HI\ gas in four Virgo spiral galaxies.  
Three color optical stellar images of our galaxy sample from SDSS overlaid with \HI\ contours from
VIVA \citep{chung09}. Unlike tidal stripping, ram pressure stripping has not
disturbed the stellar disks of our galaxy sample. The scalebars in the bottom-left of each
panel represents one arcminute.}
\label{sdss}
\end{figure}

Figure~\ref{sdss} shows three-color optical images for our sample of galaxies from the Sloan Digital Sky 
Survey (SDSS) overlaid with contours of \HI\ from the VLA Imaging of Virgo in Atomic
Hydrogen survey \citep[VIVA; ][]{chung09}.  The \HI\ gas morphology in each of these galaxies appears to be
pushed from one side and lop-sided; while the optical stellar disk appears  unaffected.

%A comparison of dynamical models with observations concluded that NGC 4330 and NGC 4501 have not reached
%the peak ram pressure; while NGC 4522 is currently near peak pressure \citep{vollmer09}.  This is consistent
%with stellar population modelling by \citet{crowl08} who found that star formation quenching timescales
%are correlated to the time since peak ram pressure.  The quenching timescales of NGC 4402 and NGC 4522 are
%estimated to be $<50$ Myr and 200 Myr, respectively \citep{crowl08}.

%Strong local radio deficits have been observed in NGC 4330, NGC 4402 and NGC 4522 in the sample of Virgo 
%spirals studied by \citet{murphy09}, suggesting that these galaxies are currently experiencing strong
% ram pressure.  

All four spirals are approximately 1.5--3.3 degrees ($\sim$0.4--0.9 Mpc) from the cluster 
center where many galaxies exhibit evidence for ongoing ram pressure stripping. It should be noted that  
NGC 4330, NGC 4402 and NGC 4522  are small spirals ($\sim0.2-0.3$ $L_*$)%\footnote{$L_*$ is the characteristic
%luminosity which defines the divide between the low- and high-luminosity galaxies and describes the dominant
%contribution towards the overall luminosity density \citep{schechter76}}) 
 which are highly-inclined, while, NGC 4501 is a moderately-inclined massive spiral ($\sim2$ $L_*$).  
Our sample favors highly-inclined galaxies because in such galaxies, extraplanar features (e.g.\ gas tails 
resulting from ram pressure stripping) are more easily identified and distinguished from the main 
disk of the galaxy. Table~\ref{samprops} lists the properties of our sample galaxies.

\begin{table*}
\small{
\begin{center}
\caption{Properties of our galaxy sample}
\vspace{0.2cm}
\label{samprops}
\begin{tabular}{ccccccccc}
\tableline
\tableline
 & $\alpha$& $\delta$ &$v_{\rm{HI}}$&Inclination & $D_{\rm{M87}}$ & $B_{T}$& $def_{\rm{HI}}$ & log $\frac{M_{\rm{HI}}}{L_K}$ \\
\multicolumn{1}{c}{Galaxy} & (J2000) & (J2000) & (\kms) & ($^{\circ}$) &($^{\circ}$) &(mag) & & ($M_{\odot}$/$L_{\odot}$)\\
\multicolumn{1}{c}{(1)} & (2) &(3) &(4) & (5) & (6) & (7) & (8) & (9) \\
\tableline
NGC 4330 & 12:23:16.14 & 11:22:01.5 & 1566 & 79 &2.3 &13.09 &$0.80\pm 0.04$ &$-1.17$\\
NGC 4402 & 12:26:07.71 & 13:06:50.3 & 236 &  80 &1.5 &12.55 & $0.74\pm 0.12$ &$-1.76$\\
NGC 4501 & 12:31:58.79 & 14:25:05.0 & 2278 & 62 & 3.0 & 10.36 & $0.58\pm 0.12$ &$-1.97$\\
NGC 4522 & 12:33:39.44 & 09:10:30.7 & 2331 & 79 &3.3 &12.99 &$0.86\pm 0.02$ &$-1.22$\\
\tableline 
\end{tabular}
\tablecomments{Col.\ (1): Name of galaxy. Col.\ (2): Right ascension.  Col.\ (3): Declination.  
Col.\ (4): \HI\ velocity \citep{chung09}.  Col.\ (5): Inclination from \citet{koopmann01} or \citet{wong10a}. 
Col.\ (6): Distance from M87 \citep{koopmann01}.  Col.\ (7): Total $B$ magnitude from
the RC3 catalog  \citep{devaucouleurs91}.   Col.\ (8): \HI\ deficiency as defined by 
$def_{\rm{\sc{Hi}}}\,=\, 0.37\, -\, \rm{log}\, \bar{\Sigma}_{\rm{\sc{HI}}}$  \citep{chung09}.  
Col.\ (9): Log of \HI\ mass-to-light ratio  \citep{chung09}.  }
\end{center}}
\end{table*}

\subsection{Observations}

For the purposes of our project we are interested in studying the extranuclear and/or extraplanar regions of the
galaxies which are being most affected by the ICM--ISM interaction.  Therefore we did not perform spectral
mapping of entire galaxies but rather, deeper observations of individual region(s) within each galaxy.

We target the leading edges of the ICM--ISM interactions in each of our sample galaxies (N4330, N4402\_1, N4501\_SW 
\& N4522\_NE).  In the three most highly-inclined galaxies (NGC 4330, NGC 4402 and NGC 4522), our 
observing apertures include the outer edges of the truncated gas disks as well as the extraplanar regions above 
or below the disks. For comparison, we also study a region dominated by the main star-forming disk of NGC 4402 (N4402\_2); 
two regions on the trailing sides of the disks of NGC 4501 and NGC 4522 (N4501\_NE \& N4522\_SW1); as well 
as an extraplanar star-forming region in NGC 4522 (N4522\_SW2). Table~\ref{targtab} lists the positions and exposure times
of the 8 targeted regions discussed in this paper.  The positions of the 6 pointings and 8  regions are as illustrated 
in panels (a) and (b) of Figure~\ref{map_panel}.

\begin{table}
\scriptsize{
\begin{center}
\caption{Details of our observations}
\vspace{0.2cm}
\label{targtab}
\begin{tabular}{lccccc}
\tableline
\tableline
 & RA & Dec & Offset & PA &Exp. time \\
\multicolumn{1}{c}{Target} & (J2000)&(J2000)&(\arcmin) & ($^{\circ}$) & (hours)\\
\multicolumn{1}{c}{(1)} & (2) &(3) &(4) & (5) & (6)\\
\tableline
N4330         & 12:23:20.5 & 11:22:33 & 1.19 & 88.1 & 2.7\\
N4402\_1        & 12:26:10.0 & 13:06:39 & 0.59 &91.3 & 3.4\\
N4402\_2      & 12:26:09.7 & 13:06:49  & 0.52  & 84.5  &3.4\\
N4501\_SW & 12:31:56.6 & 14:24:31 & 0.78& $-94.1$&2.4\\
N4501\_NE & 12:32:03.0 & 14:25:57 & 1.34 & 86.7 &2.4\\
N4522\_NE & 12:33:41.0 & 09:10:56 & 0.57 & 85.8&3.1\\
N4522\_SW1 & 12:33:37.0 & 09:10:16 & 0.65 & $-91.5$ &3.8\\
N4522\_SW2 & 12:33:37.0 & 09:10:08 & 0.77 & $-118.1$ &3.8\\
\tableline 
\end{tabular}
\tablenotetext{}{Col.\ (1): Identification of our targets. The names are abbreviated forms
of their NGC classifications.  Col.\ (2): Right ascension center of our targeted regions.  Col.\ (3): Declination
of our targeted regions.  Col.\ (4): Distance offset from the galaxy centers.  Col.\ (5): Position angle 
offset of target centers (rotating from the North to the East).  Col.\ (6): Total exposure time. }
\end{center}}
\end{table}

%N4330 & 12:23:20.5 & 11:22:33 & 2.34$\times$1.77 & 0.76$\times$0.81 & 9522\\
%N4402 & 12:26:10.0 & 13:06:39 & 2.24$\times$1.94 &0.86$\times$0.76 &12147\\
%N4501\_SW & 12:31:56.6 & 14:24:31 & 2.26$\times$1.91 & 0.74$\times$0.82&8596\\
%N4501\_NE & 12:32:03.0 & 14:25:57 & 2.26$\times$1.91 & 0.74$\times$0.82 &8594\\
%N4522\_NE & 12:33:41.0 & 09:10:56 & 2.29$\times$1.83 & 0.81$\times$0.79&11142\\
%N4522\_SW & 12:33:37.0 & 09:10:16 & 2.28$\times$1.84 & 0.92$\times$0.73 &13701\\

\begin{table}
\footnotesize{
\begin{center}
\caption{Properties of the IRS SL and LL modules \citep{houck04}.}
\vspace{0.2cm}
\label{techtab}
\begin{tabular}{lcccc}
\tableline
\tableline
Properties & SL2 & SL1 & LL2 & LL1 \\
\tableline
Pixel scale & 1.8\arcsec\ & 1.8\arcsec\ & 5.1\arcsec\ & 5.1\arcsec\ \\
Spectral resolution & 60--120 &60--120 &60--120 &60--120 \\
$\lambda$ ($\mu$m) & 5.2--7.7 & 7.4--14.5 & 14.0--21.3 & 19.5--38.0\\
Slit size (\arcsec)&3.6$\times$57 &  3.7$\times$57 & 10.5$\times$168 &10.7$\times$168 \\
\tableline
\end{tabular}
\end{center}}
\end{table}

Our sample was observed in July 2008 using the IRS instrument on board the {\it{Spitzer}} Space Telescope
 \citep{houck04}. 
Spectral mappings within the wavelength range of 5 to 38 $\mu$m were completed using the Short-Low (SL) 
and Long-Low  (LL) modules. Each module comprises of 2 sub-modules for the first-order and second-order spectra. 
Table~\ref{techtab} lists the properties of the SL and LL modules.

We modeled our observation strategy loosely upon the spectral mapping performed by the SINGS Legacy team
\citep{kennicutt03} with the exception that we did not map the cores of galaxies.  As such, our 
integration times are longer as we need to probe to fainter surface brightness levels at the outer edges of  
galaxies.  To reduce the effects of bad pixels, each perpendicular slit pointing is 
offset by half the slit width.  We then employed two cycles for each observation to provide greater 
redundancy against rogue pixel identification and cosmic ray detections.

Within the wavelength range of 5 to 38 $\mu$m, we expect to find three types of emission features in
addition to the MIR dust continuum, which is produced by a combination of transiently heated 
grains at various temperatures and a hot thermal dust component.
 These emission features include: (1)  emission from PAH molecules.  Most of the PAH emission  can be 
loosely grouped into four large emission complexes around the wavelengths of 7.7, 11.3, 12.7 and 
17 $\mu$m;  %In addition to the PAH emission lines, the spectrum between 10 $\mu$m to 20 $\mu$m also 
%potentially includes absorption (e.g.\ silicate absorption between 7.7 and 11.3 $\mu$m) and emission 
%features (at 12.0, 12.6, 13.6, 14.3, 15.9, 16.4, 18.9 $\mu$m) which may appear blended with the PAH and 
%\htwo\ transitional and fine structure lines.  Beyond 20 $\mu$m, a broad dust feature can be observed 
%centered around 33.1 $\mu$m.  This particular dust feature has been observed in a sample of nearby galaxies
%\citep[SINGS; ][]{smith07b}.  
(2) emission from the rotational transitions of the ground vibrational states of warm \htwo;  %Table~\ref{h2props} lists the seven rotational transitions which would 
%be observable within our observed wavelength range as well as the upper level energies and the transition probabilities 
%of each of these transitions.   
(3) atomic fine structure line emission from various species such as 
$[$Ne {\sc{ii}}$]$, $[$Ne {\sc{iii}}$]$, $[$Si {\sc{ii}}$]$ and $[$S {\sc{iii}}$]$ which can be used
to describe the hardness of the radiation field and the electron density of our observed regions. %The ratios of these
% non-hydrogen emission lines can help distinguish between emission dominated by active galactic nuclei (AGN) 
%and that by hot young stars.  
%A summary of the most common atomic fine structure lines and their properties is as listed in Table~\ref{othemlines}.

\subsection{Data processing}

The final spectral maps  were derived from the Basic Calibrated Data (BCD) datasets generated by the 
Spitzer Science Center (SSC) pipeline (version S18.7).  Using {\sc{Cubism}} \citep[version 1.6; ][]{smith07a}, the datasets 
were further processed and the spectral maps of each pointing were generated for each of the four IRS modules.  The 
spectra  of the four IRS modules (derived from the overlap region between the four modules) were merged to produce a single 
continuous spectrum spanning 5--38 $\mu$m for each of the observed pointings.  For each pointing,  we extract a 
region of approximately 1 kpc$^2$ (approximately 0.5 arcmin$^2$).  Due to slight mismatches between the different
 spectral segments which result from small residual photometric and astrometric uncertainties, we scaled the spectral
 segments according to the method outlined by \citet{smith07b}.  The SL2 and LL2 modules are scaled to match the
 spectra from SL1 and LL1, respectively. Subsequently, the combined SL spectrum is scaled  to match the LL
 spectrum of each pointing.

To determine the accuracy of the resultant spectra, we compared the integrated flux filtered through the 8 $\mu$m IRAC 
and the 24 $\mu$m MIPS bandpasses  to that measured from the 8 $\mu$m IRAC and 24 $\mu$m MIPS images of the same 
region from our MIR imaging counterpart survey, the Spitzer Survey of Virgo \citep[SPITSOV; ][]{wong10a}.  The 8 $\mu$m
 and 24 $\mu$m measurements from the SPITSOV and IRS observations  agree to within the calibration uncertainties
 ($\sim$15\%) of each instrument. 

%agree to within 15\% of each other. Repeating the same process for the MIPS 24 $\mu$m bandpass, the average agreement between the two measurements were between 15--20\% of each other.

\section{Results}

\subsection{MIR spectra}
The emission line identification and decomposition of the calibrated spectra were performed using  
{\sc{Pahfit}}, an IDL tool for decomposing {\it{Spitzer}} IRS spectra of PAH emission sources and the recovery of weak, 
blended dust emission features including silicon absorption \citep{smith07b}.  We fit our spectra with
the default parameter settings while omitting a fit for a Galactic dust extinction model. It should be noted that our
 main results do not change had we opted to fit an extinction model  for each target.

\begin{figure}
\begin{center}
\begin{tabular}{c}
\includegraphics[scale=0.4]{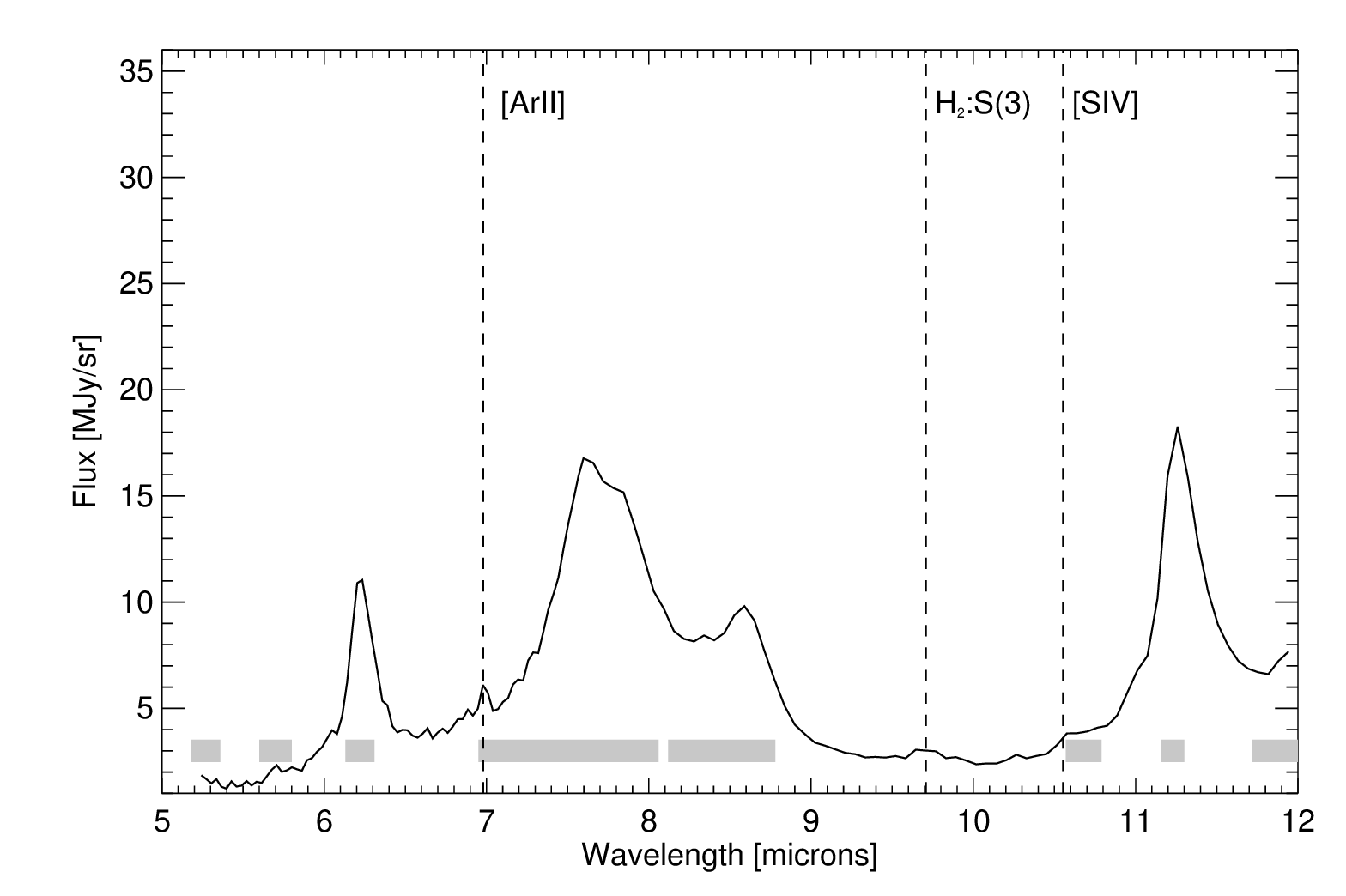}\\
\includegraphics[scale=0.4]{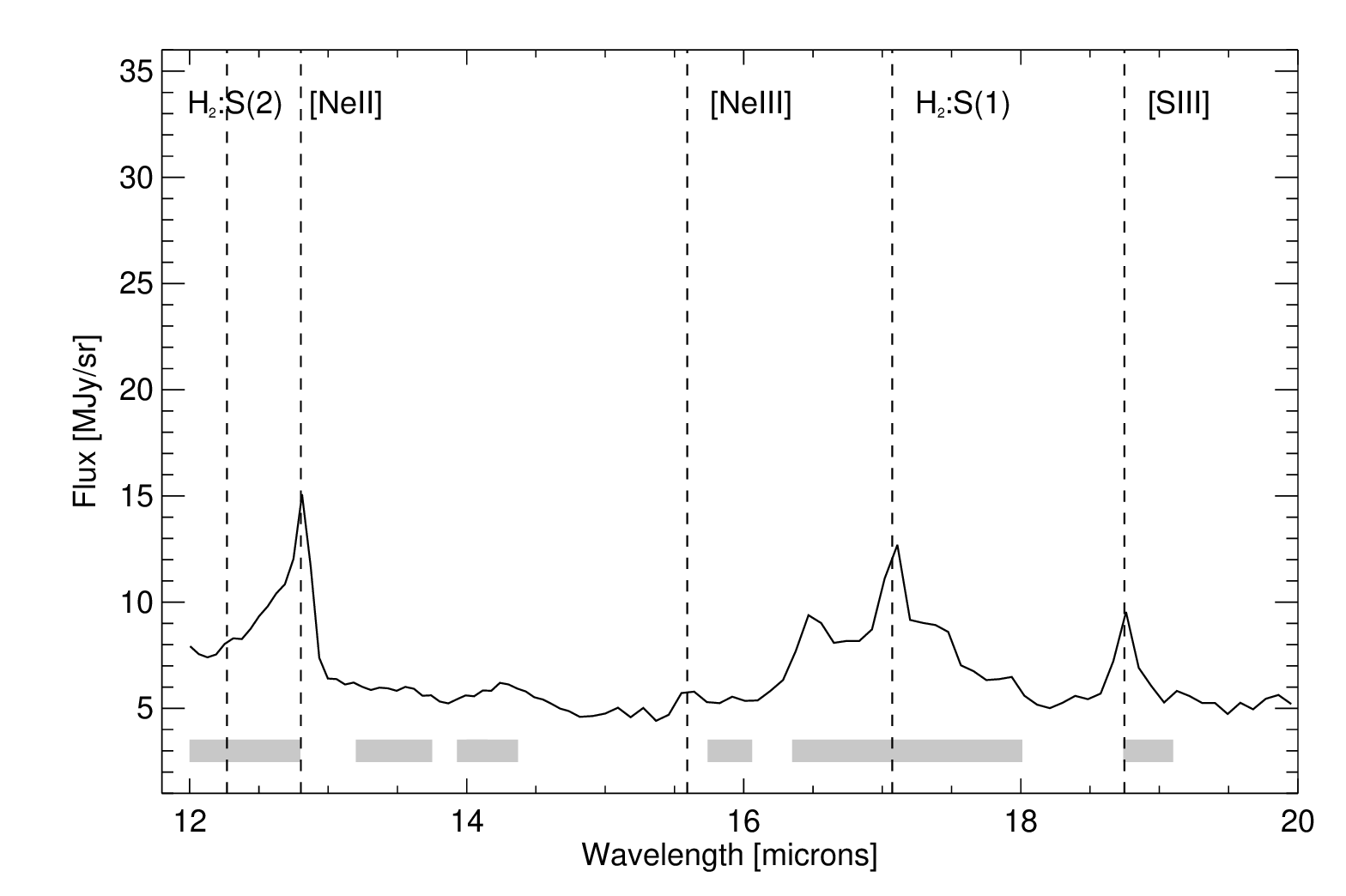}\\
\includegraphics[scale=0.4]{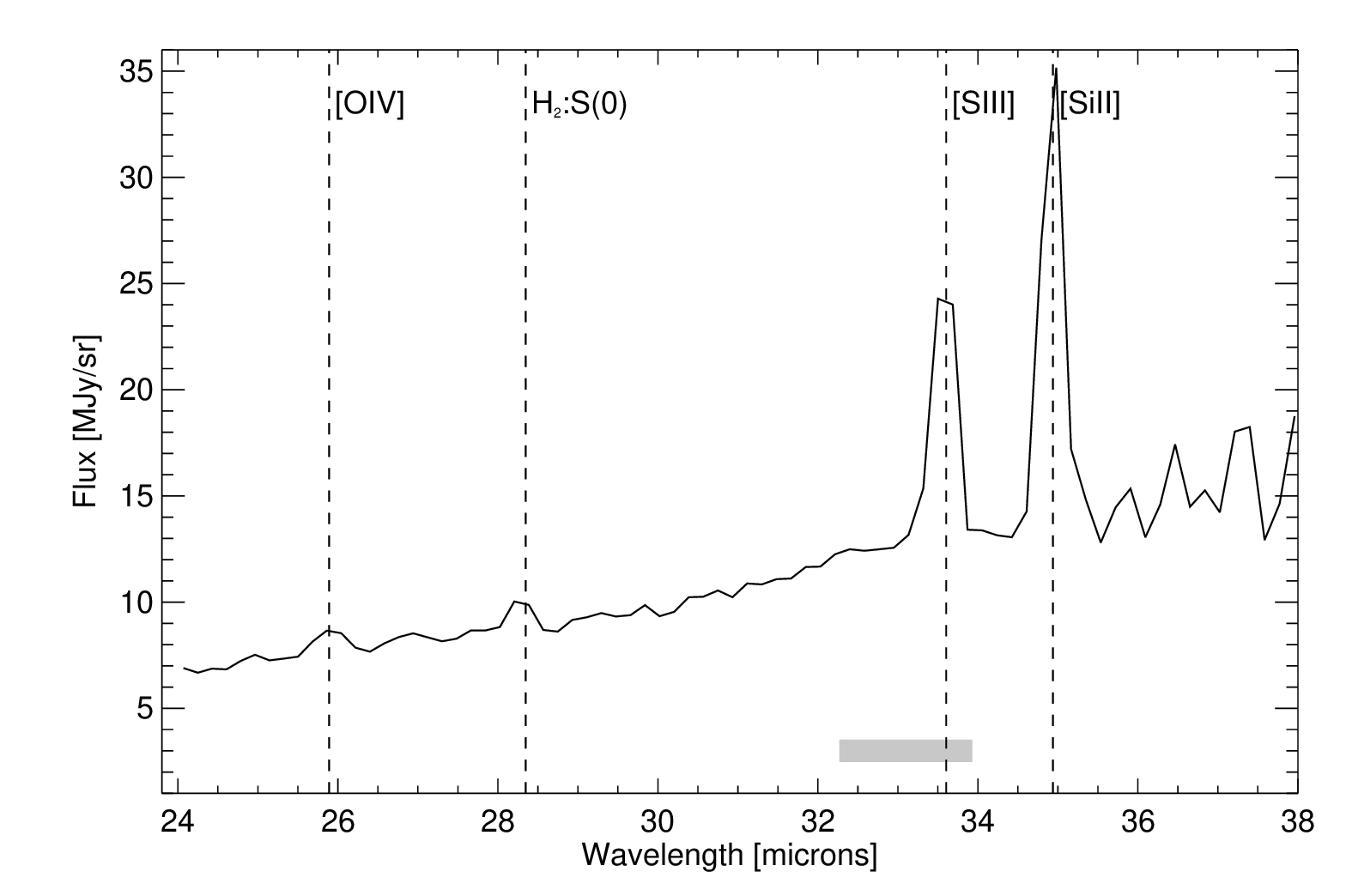}\\
\end{tabular}
\end{center}
\caption{Example IRS spectra of N4402\_1.  Fine structure atomic emission lines and rotational lines of \htwo\ are indicated
 by the labels. Shaded bars represent the wavelength range where dust features (such as PAH and silicates) are found.}
\label{egn4402}
\end{figure}

In Figure~\ref{egn4402}, we show the spectrum of N4402\_1 overlaid with labels as an example of the resultant emission lines 
found by {\sc{Pahfit}}.  The rotational \htwo\ emission and the atomic fine structure lines are marked by vertical 
dashed lines.  The wavelength range at which emission from dust features are observed are defined by the span of the
 shaded horizontal gray bars. As our target regions include star-forming regions, the resulting PAH emission 
complexes at the shorter wavelengths ($<10$ $\mu$m) are overwhelmingly stronger than the higher excitation 
\htwo\ rotational transitions that are at coincident wavelengths.

The emission line intensities and uncertainties determined by \pahfit\ are listed in Table~\ref{obstab}.  
The \htwo\ rotational lines are given at the beginning of the table, followed by the fine-structure lines and then 
the emission from the PAH dust complexes. Upper limits are provided where the signal-to-noise ratio is less than 3.

{
\begin{landscape}
\begin{table*}
\scriptsize{
\begin{center}
\caption{Integrated emission line intensities of our spectra as measured by \pahfit.}
\vspace{0.2cm}
\label{obstab}
\begin{tabular}{lcccccccc}
\tableline
\tableline
&  & &Object & & & \\
Emission line feature & N4330 & N4402\_1 & N4402\_2 &N4501\_SW & N4501\_NE & N4522\_NE &N4522\_SW1 & N4522\_SW2\\
 & (1) & (2) & (3) & (4) &(5) &(6) &(7)\\
\tableline
\multicolumn{8}{c}{\em{Rotational \htwo\ emission}} \\
\\
%%$S(0)$ & 4.6 $\pm1.2$ & 17.8 $\pm1.6$ & 23.3 $\pm1.6$ &10.4 $\pm1.3$ & 4.3 $\pm1.1$ & 8.0 $\pm1.2$ & 4.1 $\pm1.1$\\
$S(0)_{28.2 \mu m}$ & 0.17 $\pm0.04$ & 0.48 $\pm0.04$& 0.63 $\pm$0.04&0.42 $\pm$0.05&0.19 $\pm$0.05& 0.26$\pm$0.04&0.14 $\pm$0.04& 0.61 $\pm$0.08\\
%%$S(1)$ & 8.7 $\pm2.4$ & 64.6 $\pm3.3$ & 90.5 $\pm3.4$ &32.0 $\pm2.3$ & $<$6.5     & 18.7 $\pm2.3$ & 8.5 $\pm2.4$\\
$S(1)_{17.0 \mu m}$ & 0.31  $\pm0.09$& 1.74 $\pm$0.09& 2.44 $\pm0.09$&1.28 $\pm$0.09&$<$0.29     & 0.62 $\pm$0.08& 0.29 $\pm$0.08& 0.55 $\pm$0.13\\
%%$S(2)$ & 6.3 $\pm2.0$ & 25.2 $\pm2.5$ & 33.7 $\pm1.9$ &$<$20.4      & $<$7.0     & 5.2 $\pm1.3$ & $<$5.2\\
$S(2)_{12.3 \mu m}$ & 0.23 $\pm0.07$ & 0.68 $\pm$0.07& 0.91 $\pm$0.05&$<$0.82      &$<$0.32     & 0.17 $\pm$0.04&$<$0.18 &$<$0.62 \\
%%$S(3)$ & $<$6.5       &16.7 $\pm2.0$  &  33.6 $\pm2.3$& $<$14.8     &$<$15.5    &$<$6.7 & $<$8.6\\
$S(3)_{9.7 \mu m}$ & $<$0.23  & 0.45 $\pm$0.05& 0.91 $\pm$0.06      & $<$0.59     &$<$0.70    &$<$0.22& $<$0.29 &$<$0.45  \\
%%$S(4)$ & $<$19.3      & 45.6$\pm4.6$  & 23.5 $\pm5.2$& $<$12.1     & $<$12.1    &$<$35.7 &$<$12.7\\
$S(4)_{8.0 \mu m}$ & $<$0.69  & $<$0.36 & $<$0.42   & $<$ 0.48    &$<$0.54     &$<$1.18 &$<$0.43&$<$0.89\\
%$S(5)$ & --- & $<$25.7 &$<$26.0 & $<$14.3 & ---&---&$<$13.0\\
%$S(6)$ & $<$2.4 & --- & ---& --- & --- &---&$<$1.6\\
%$S(7)$ & $<$2.0 & --- & ---& $<$11.3 & --- & $<$7.1 &$<$6.2\\
\tableline
\multicolumn{8}{c}{\em{Atomic fine structure emission}}  \\
\\
%%$[$Ar \sc{ii}$]$ &$<$26.3 & 59.3 $\pm$6.5 &83.3 $\pm$7.9 & $<$28.4 & --- &$<$19.3 &$<$20.8\\
$[$Ar \sc{ii}$]_{7.0 \mu m}$   &$<$0.95 &1.60 $\pm$0.18&2.25 $\pm$0.21 & $<$1.14 & --- &$<$0.64 &$<$0.71& ---\\
%%$[$Ar \sc{iii}$]$ & ---       & --- &$<$14.0& --- &--- & $<$7.1 & ---\\
$[$Ar \sc{iii}$]_{9.0 \mu m}$ & ---       &  --- &$<$0.38 & --- &--- &$<$0.23 & ---& ---\\
%%$[$S \sc{iv}$]$ &$<$5.5 & $<$8.5 &19.5 $\pm5.5$ & $<$9.0& --- &$<$9.9 &$<$6.2 \\
$[$S \sc{iv}$]_{10.5 \mu m}$   &$<$0.20 &$<$0.23&0.53 $\pm$0.15&$<$0.36& --- &$<$0.33&$<$0.21&$<$0.38\\
%%$[$Ne \sc{ii}$]$ &21.5 $\pm$1.6 & 148 $\pm2$ & 208 $\pm2$        & 59.8 $\pm2.8$ &5.7 $\pm1.4$  &61.2 $\pm2.2$ &9.1 $\pm2.0$\\
$[$Ne \sc{ii}$]_{12.8 \mu m}$ & 0.77 $\pm$0.06 & 4.00 $\pm$0.05 & 5.62 $\pm$0.05& 2.39 $\pm$0.11&0.26 $\pm$0.06&2.02 $\pm$0.07&0.31 $\pm$0.07&$<$0.52\\
%%$[$Ne \sc{iii}$]$ &$<$4.9 &21.2 $\pm3.4$ &32.2 $\pm2.4$ &18.0 $\pm2.2$ &5.2 $\pm1.6$  & 8.0 $\pm1.8$ &$<$6.2\\
$[$Ne \sc{iii}$]_{15.6 \mu m}$ &$<$0.18& 0.57 $\pm$0.09 &0.87 $\pm$0.06&0.72 $\pm$0.09&0.23 $\pm$0.07&0.26 $\pm$0.06&$<$0.21&$<$0.38\\
%%$[$S \sc{iii}$]$$_{18\mu m}$ & 10.0 $\pm1.7$ & 55.8 $\pm2.4$ &69.0 $\pm2.4$ & 29.2 $\pm2.1$& 6.0 $\pm1.3$ & 21.8 $\pm1.7$ &$<$5.2\\
$[$S \sc{iii}$]$$_{18\mu m}$ & 0.36 $\pm$0.06 & 1.51 $\pm$0.06  &2.76 $\pm$0.10&1.31 $\pm$0.09&0.27 $\pm$0.06& 0.72 $\pm$0.06&$<$0.18&$<$0.41\\
%%$[$O \sc{iv}$]$ & --- & 24.3 $\pm2.8$ &12.7 $\pm3.0$ &$<$15.2 &$<$4.7 & --- &---\\
$[$O \sc{iv}$]_{26.9 \mu m}$ & --- & 0.66 $\pm$0.08 & 0.34 $\pm$0.08&$<$0.61 &$<$0.21& ---&---&---\\
%%$[$Fe \sc{ii}$]$ &$<$4.1 & $<$7.4 & 17.7 $\pm2.9$ &$<$11.4 &$<$4.7 &6.1 $\pm1.5$  & --- \\
$[$Fe \sc{ii}$]_{27.0 \mu m}$ &$<$0.15 & $<$0.20 & 0.48 $\pm$0.08&$<$0.46 &$<$0.21&0.20 $\pm$0.05& ---&$<$0.33\\
%%$[$S \sc{iii}$]$$_{33\mu m}$ & 13.4 $\pm2.3$ & 124 $\pm3$ & 131 $\pm3$     &31.9 $\pm2.6$ & $<$4.4 & 29.8 $\pm2.0$& $<$6.5 \\
$[$S \sc{iii}$]$$_{33\mu m}$ & 0.48 $\pm0.08$ & 3.35 $\pm$0.08& 3.54 $\pm$0.08&1.28 $\pm$0.10& $<$0.20&0.98 $\pm$0.07&$<$0.22& 0.54 $\pm$0.12\\
%%$[$Si \sc{ii}$]$ & 14.3 $\pm2.4$ & 178 $\pm3$ &236 $\pm3$      &57.2 $\pm2.7$ & 8.1 $\pm1.7$ &38.1 $\pm4.7$  &14.8 $\pm2.7$\\
$[$Si \sc{ii}$]_{34.8 \mu m}$ & 0.51 $\pm$0.09& 4.81 $\pm$0.08& 6.37 $\pm$0.08&2.29 $\pm$0.11&0.36 $\pm$0.08& 1.26 $\pm$0.16&0.50 $\pm$0.09& 0.88 $\pm$0.23\\
\tableline 
\multicolumn{8}{c}{\em{PAH emission features and complexes}}  \\
\\
%%7.7 $\mu$m & 1.13$(\pm 0.07)\times 10^3$ &7.96$(\pm 0.07)\times 10^3$&   1.25$(\pm 0.01)\times 10^4$&   2.86$(\pm 0.06)\times 10^3$& 5.39$(\pm 0.43)\times 10^2$&2.48$(\pm 0.06)\times 10^3$ &4.59$(\pm 0.53)\times 10^2$\\
6.2 $\mu$m & 10.9 $\pm$1.0&57.2 $\pm$0.5&92.6 $\pm$0.5&32.6 $\pm$0.7&13.8 $\pm$0.6&26.9 $\pm$0.6& 6.2 $\pm$0.7& 4.2 $\pm$1.1\\
7.7 $\mu$m & 40.7 $\pm$2.5 & 215 $\pm$2 &337 $\pm$3 &114 $\pm$2& 24.3 $\pm$1.9 & 81.8 $\pm$2.0 &15.6 $\pm$1.8&12.3 $\pm$1.3\\
%11.3 $\mu$m &2.32$(\pm 0.06)\times 10^2$&1.62$(\pm 0.01)\times 10^3$ &2.62$(\pm 0.01)\times 10^3$ &7.42$(\pm 0.05)\times 10^2$&1.44$(\pm 0.06)\times 10^2$&6.14$(\pm 0.05)\times 10^2$ &1.13$(\pm 0.06)\times 10^2$\\
11.3 $\mu$m &8.4$\pm$0.2&43.7 $\pm$0.3 & 70.7 $\pm$0.3 & 29.7 $\pm$0.2 &6.5 $\pm$0.3& 20.3 $\pm$0.2& 3.8 $\pm$0.2& 5.6 $\pm$0.4\\
%12.6 $\mu$m &89.4 $\pm$8.5 &1.00$(\pm 0.01)\times 10^3$ &1.61$(\pm 0.01)\times 10^3$ &3.53$(\pm 0.11)\times 10^2$&59.9 $\pm$7.4 &3.18$(\pm 0.08)\times 10^2$ &61.2 $\pm$7.1\\
12.6 $\mu$m &3.2 $\pm$0.3 &27.0 $\pm$0.3&43.5 $\pm$0.3&14.1 $\pm$0.4&2.7 $\pm$0.3 &10.5 $\pm$0.3 &2.1 $\pm$0.2 &2.6 $\pm$0.5 \\
%17.0 $\mu$m &51.7 $\pm$12.5 &8.70$(\pm 0.18)\times 10^2$ &1.19$(\pm 0.02)\times 10^3$ &4.71$(\pm 0.15)\times 10^2$&54.8 $\pm$9.0 &2.13$(\pm 0.14)\times 10^2$ &$<$28.0 \\
17.0 $\mu$m &1.86$\pm$0.45 &23.5 $\pm$0.5&32.1 $\pm$0.5 &18.8 $\pm$0.6 &2.47 $\pm$0.41 &7.03 $\pm$0.46 &$<$0.95&3.0 $\pm$0.6\\
\tableline 
\end{tabular}
\tablecomments{Our observed regions are listed from Columns (1) to (8) and the solid angle
over which these flux densities were integrated are $3.6\times 10^{-9}$, $2.7\times 10^{-9}$, $2.7\times 10^{-9}$, 
$4.0\times 10^{-9}$, $4.5\times 10^{-9}$, $3.3\times 10^{-9}$,  $3.4\times 10^{-9}$ and $9.96 \times 10^{-9}$ steradians, 
respectively.  The line intensities are given in units of $10^{-17}$W m$^{-2}$ .}
\end{center}}
\end{table*}
\end{landscape}}

\subsection{Warm \htwo\  emission line maps}

\begin{figure*}[!]
\begin{center}
\begin{tabular}{c}
\includegraphics[scale=.7]{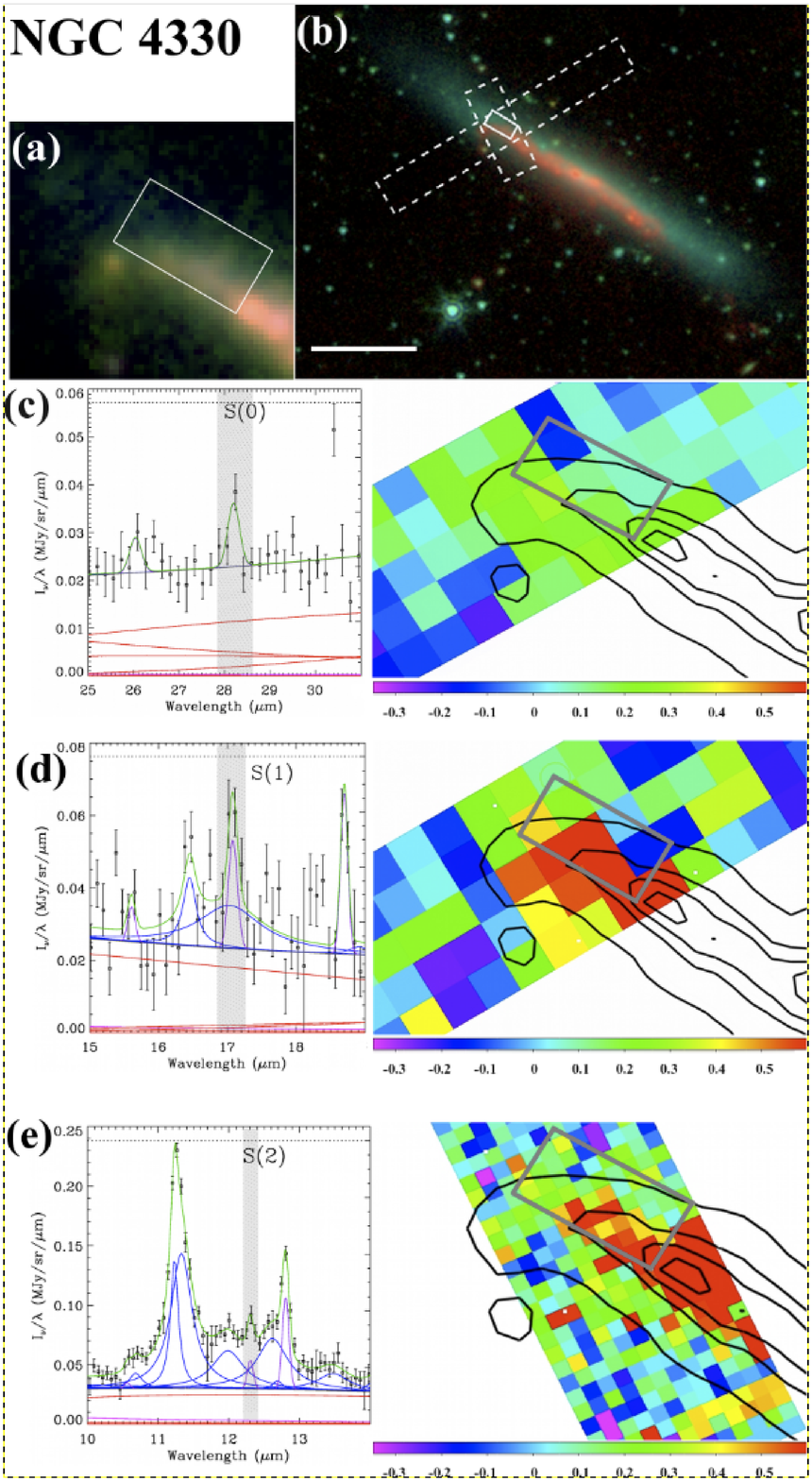}\\
%\\
%\\
\end{tabular}
\end{center}
\caption{\scriptsize{Rotational \htwo\ emission line maps of our observations (where North and East are aligned to the top and left of the page).  Panels (a) and (b) show the 3-color IRAC observations of each target galaxy. Blue, green and red represent the 3.6 $\mu$m, 5.8 $\mu$m and continuum-subtracted 8 $\mu$m PAH emission in panel (a) and 3.6 $\mu$m, 4.5 $\mu$m and continuum-subtracted 8 $\mu$m PAH emission in panel (b); respectively.  The overlaid  dashed-line boxes show the field-of-view of the SL and LL modules, respectively. The solid-line box shows the region from which the spectra were extracted.  The scale bar represents 1 arcminute.  Panel (a) shows the magnified version of our regions-of-interest. Panels (c), (d) and (e) show the \htwo\ $S(0)$, $S(1)$ and $S(2)$ rotational emission line maps overlaid with contours of the 8 $\mu$m PAH map, respectively.  
%It should be noted that an estimate of the PAH contamination has been subtracted from the observed $S(1)$ and $S(2)$ maps. The observed emission line is highlighted within the gray-shaded  region of the accompanying spectrum in each panel.  
 In the accompanying spectrum, the red lines represent the thermal dust continuum at various dust temperatures.  The dust emission lines and the atomic (and molecular) lines are shown by the blue and purple solid lines and the green line shows the fit to the sum of the continuum and emission line components. }}
\label{map_panel}
\vspace{0.5cm}
\end{figure*}

Using the spectral mapping capabilities of the IRS instrument, we map the distribution of \htwo\ 
 in various rotational transitions within our sample.  The spatial distributions of the $S(0)$, $S(1)$ 
and $S(2)$ rotational transitions of \htwo\ for each of our observed regions are shown in Figure~\ref{map_panel}.  
Five panels are shown for each IRS pointing in Figure~\ref{map_panel}: Panel (a) shows a magnified view of 
the targeted region(s) in our galaxy sample. In this 3-color IRAC image, the blue, green and red represent 
the 3.6 $\mu$m, 5.8 $\mu$m and continuum-subtracted 8 $\mu$m PAH emission \citep{helou04}, respectively).  Panel (b) 
shows the 3-color IRAC image of the target  overlaid with two white dashed-lined rectangles  delineating 
the field-of-view of  the SL and LL  modules. In panel (b) the blue, green and red colors represent the 3.6, 4.5 and
 continuum-subtracted 8 $\mu$m PAH emission, respectively. The solid rectangle shows the region where  the 
spectrum is extracted.  Panels (c), (d) and (e) show the $S(0)$, $S(1)$ and $S(2)$ spectra and the emission line maps overlaid with 
contours of the  8 $\mu$m PAH map, respectively.

Observations of star-forming regions on scales of several hundred parsecs find the PAH emission to originate from
shell-like regions around the central \htwo\ regions (e.g.\ \citet{bendo08} and references therein). However, we do 
not observe any spatial differences between the peaks of our warm \htwo\ emission and that of the PAH emission due to 
the low angular resolution of our maps. %More detailed analysis of this region 
%can be found in Section~\ref{excdiag}

We find significant $S(0)$ emission not only from star-forming regions within the galaxy disks, but also from 
the south-western extraplanar star-forming region of NGC 4522 (N4522\_SW2).  The ratio between $S(0)$ and $S(1)$ 
are fairly similar in most of our target regions with the exception of N4522\_SW2.  Whereas in the disk the $S(1)$
 emission is stronger, in the extraplanar region (N4522\_SW2) the $S(0)$ emission is stronger.
The $S(0)/S(1)$ fraction of N4522\_SW2 is 2.6 times greater than the average of all the other targeted
 regions, where the mean of $S(0)/S(1)$ is approximately $0.42$ and the standard deviation is 0.15.  This result suggests that the
extraplanar gas is much colder than the gas in the disk.  While there are \HII\ regions in the stripped extraplanar
 gas, we hypothesize that the density of the heating sources in the extraplanar star-forming regions is likely 
to be lower than those in the disk regions.     Further details of our \htwo\ 
temperature estimates can be found in Section 4.3 where we determine the temperatures of the warm \htwo\ observed 
in our sampled regions using one-temperature and two-temperature dust models. 

%In addition, the observed \htwo\ in N4522\_SW2 is much colder than the \htwo\ gas in the tail of ESO 137-001, 
%possibly because ram pressure is much stronger in the Norma Cluster than in Virgo. Recent simulations 
%and studies of the Virgo cluster concluded that the general lack of X-ray
% tails in ram pressure-stripped galaxies may be an indicator of low ICM pressure \citep{tonnesen11,jachym13}.   
%In IC 3418 (a Virgo cluster galaxy with a similar star-forming tail to that in ESO 137-001), \citet{jachym13}
%found the X-ray (0.5--2 keV) luminosity limit to be 280 times weaker than that measured in ESO 137-001 \citep{sun10}.

%The estimated temperature of the warm \htwo\ component in the ESO 137-001 extraplanar tail is very similar to the 
%temperatures that we estimate for our targeted disk regions \citep{sivanandam09}. 

%Our most interesting \htwo\ maps are of NGC 4522. Although there are not significant detections of the higher order S(2) 
%emission in our regions, we do observe significant S(0) emission further away from the central disk/nuclear regions
%than our observations of the S(1) emission.  In particular, there exists a significant amount of S(0) emission in 
%the extraplanar regions on the trailing side of NGC 4522 (N4522\_SW), whereas the observed S(1) emission is found closer
%to the disk plane of the galaxy.  This result is consistent with the idea that the \htwo\ is more likely to be heated
%to higher rotational orders (also to higher temperatures) closer to the disk plane than outside the plane.

\begin{figure*}[!]
\begin{center}
\begin{tabular}{c}
\includegraphics[scale=.8]{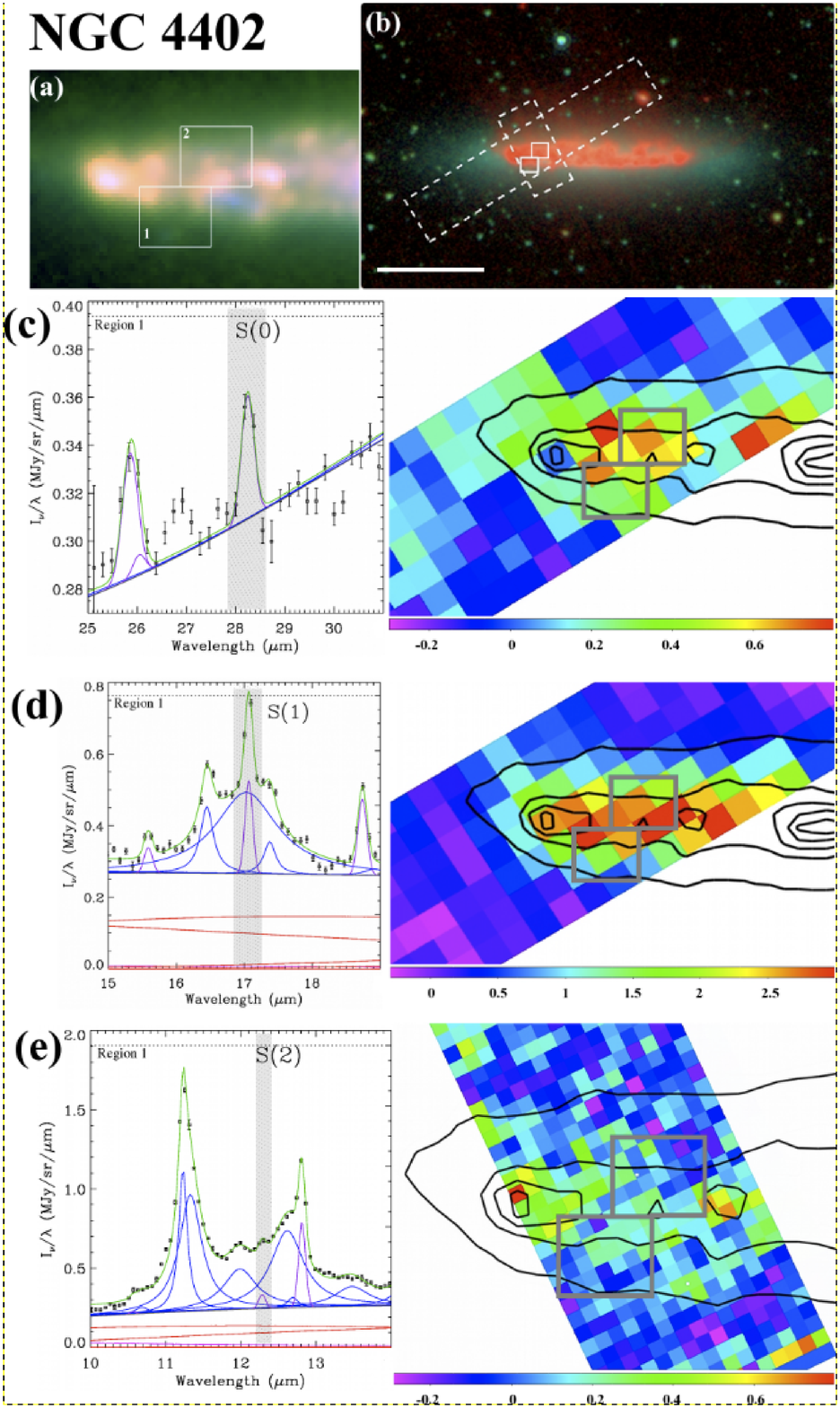}\\
\end{tabular}
\end{center}
\begin{center}
Figure~\ref{map_panel} (continued... NGC 4402).  Note that the spectra in panels (c), (d) and (e) are from Region 1. Region 1 is south-east of Region 2.
\end{center}
\end{figure*}
\begin{figure*}[!]
\begin{center}
\begin{tabular}{c}
\includegraphics[scale=.8]{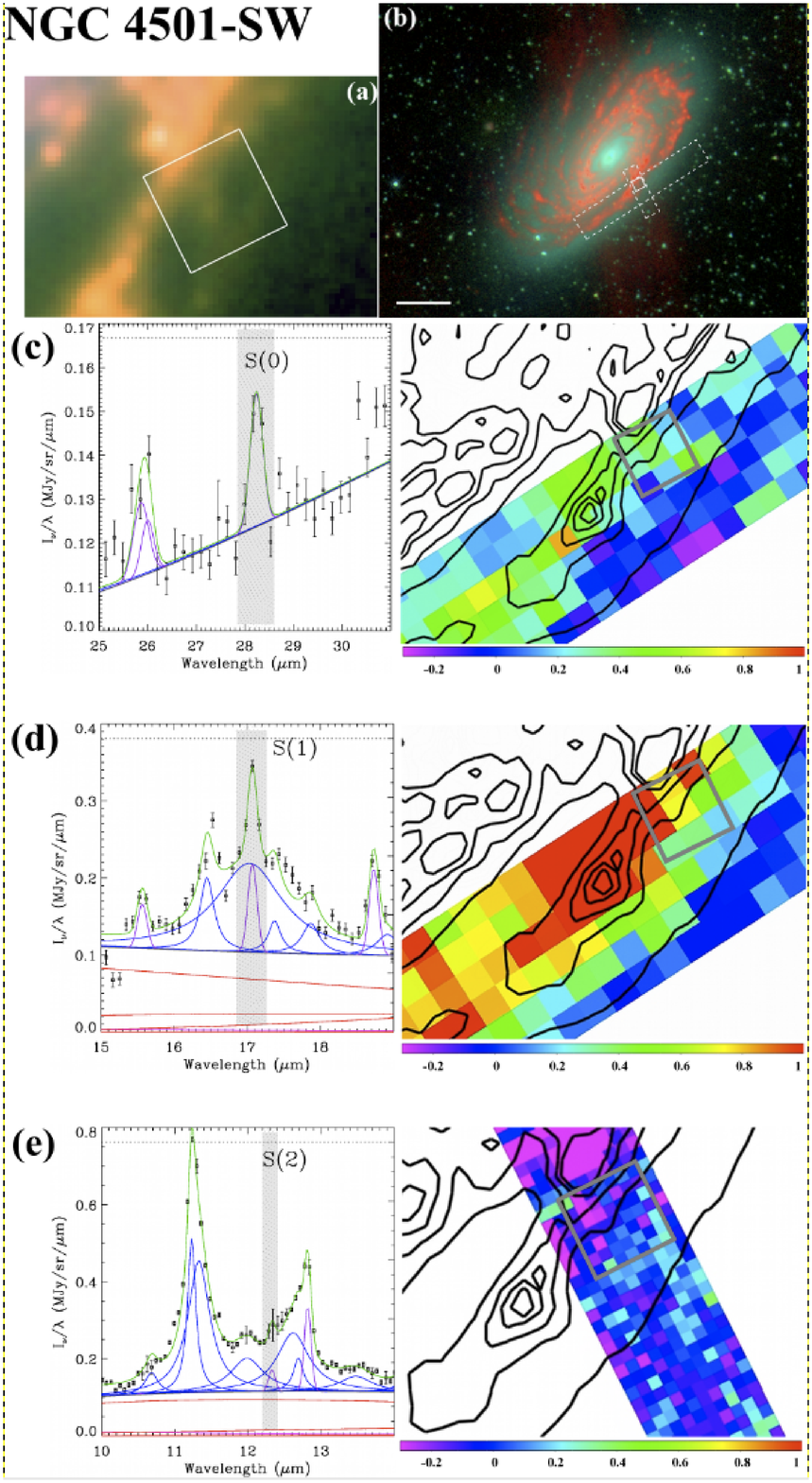}\\
\end{tabular}
\end{center}
\begin{center}
Figure~\ref{map_panel} (continued... NGC 4501 - southwest region)
\end{center}
\vspace{0.5cm}
\end{figure*}
\begin{figure*}[!]
\begin{center}
\begin{tabular}{c}
\includegraphics[scale=.8]{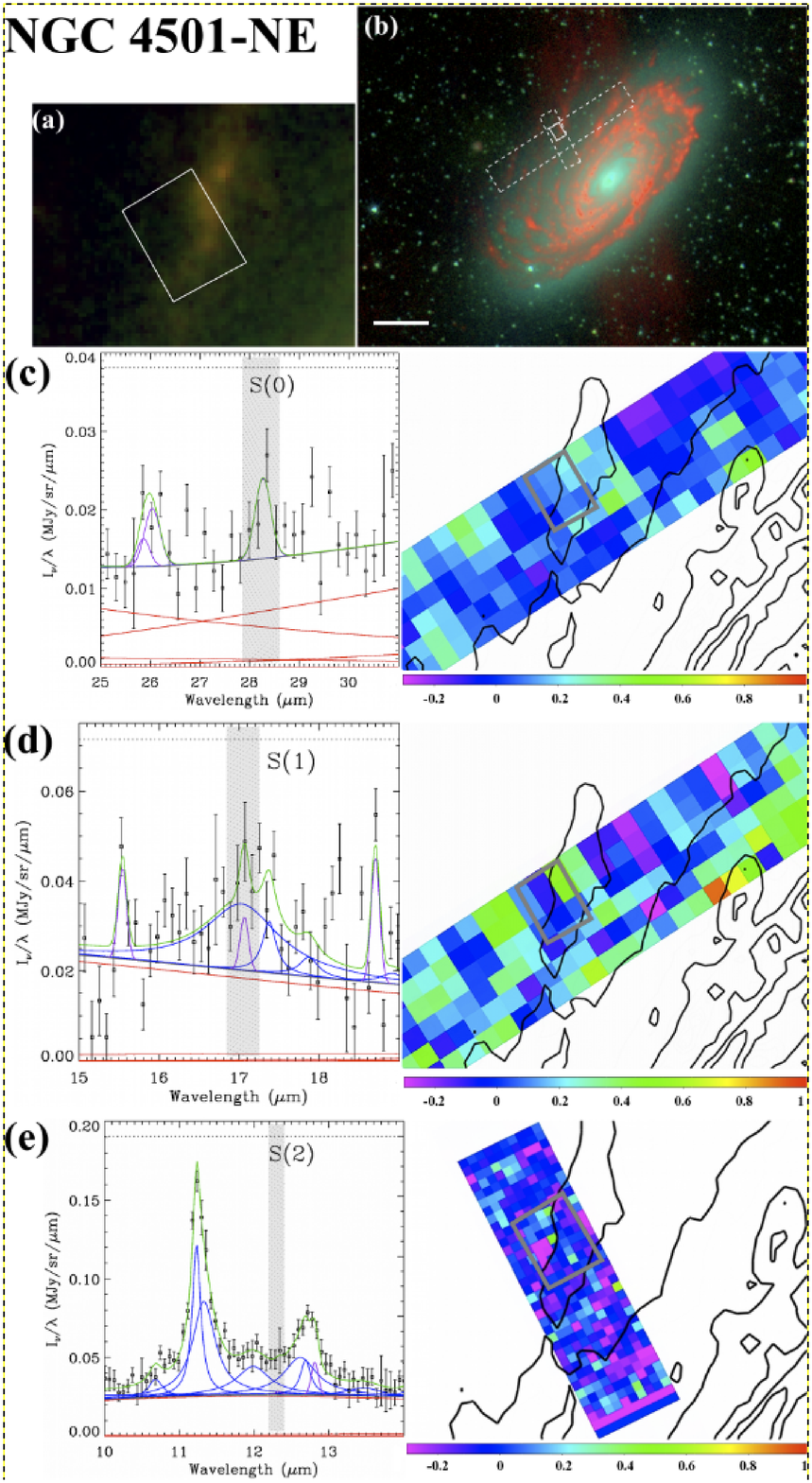}\\
\end{tabular}
\end{center}
\begin{center}
Figure~\ref{map_panel} (continued... NGC 4501 - northeast region)
\end{center}
\vspace{0.5cm}
\end{figure*}
\begin{figure*}[!]
\begin{center}
\begin{tabular}{c}
\includegraphics[scale=.8]{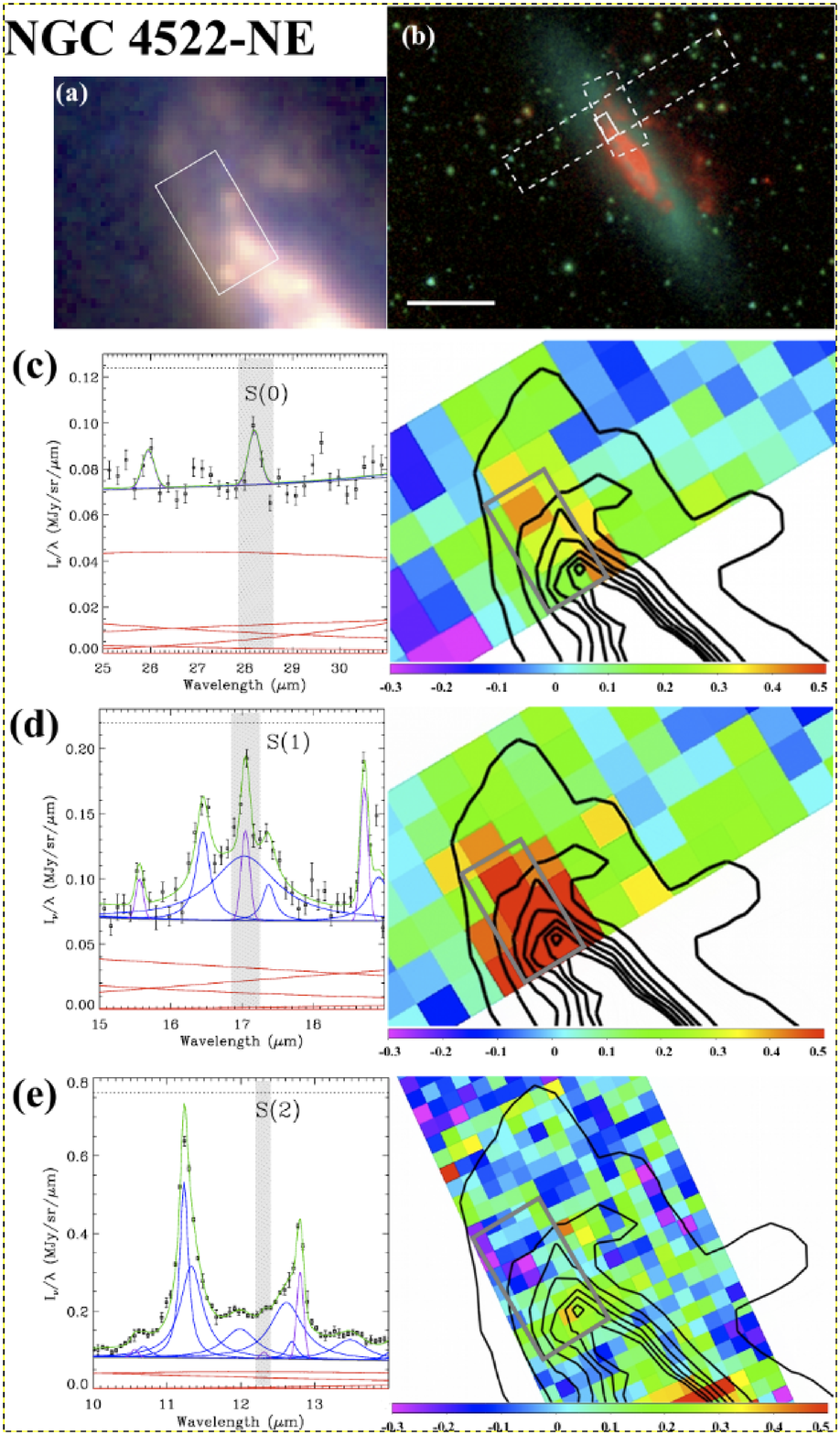}\\
\end{tabular}
\end{center}
\begin{center}
Figure~\ref{map_panel} (continued... NGC 4522 - northeast region)
\end{center}
\vspace{0.5cm}
\end{figure*}
\begin{figure*}[!]
\begin{center}
\begin{tabular}{c}
\includegraphics[scale=.8]{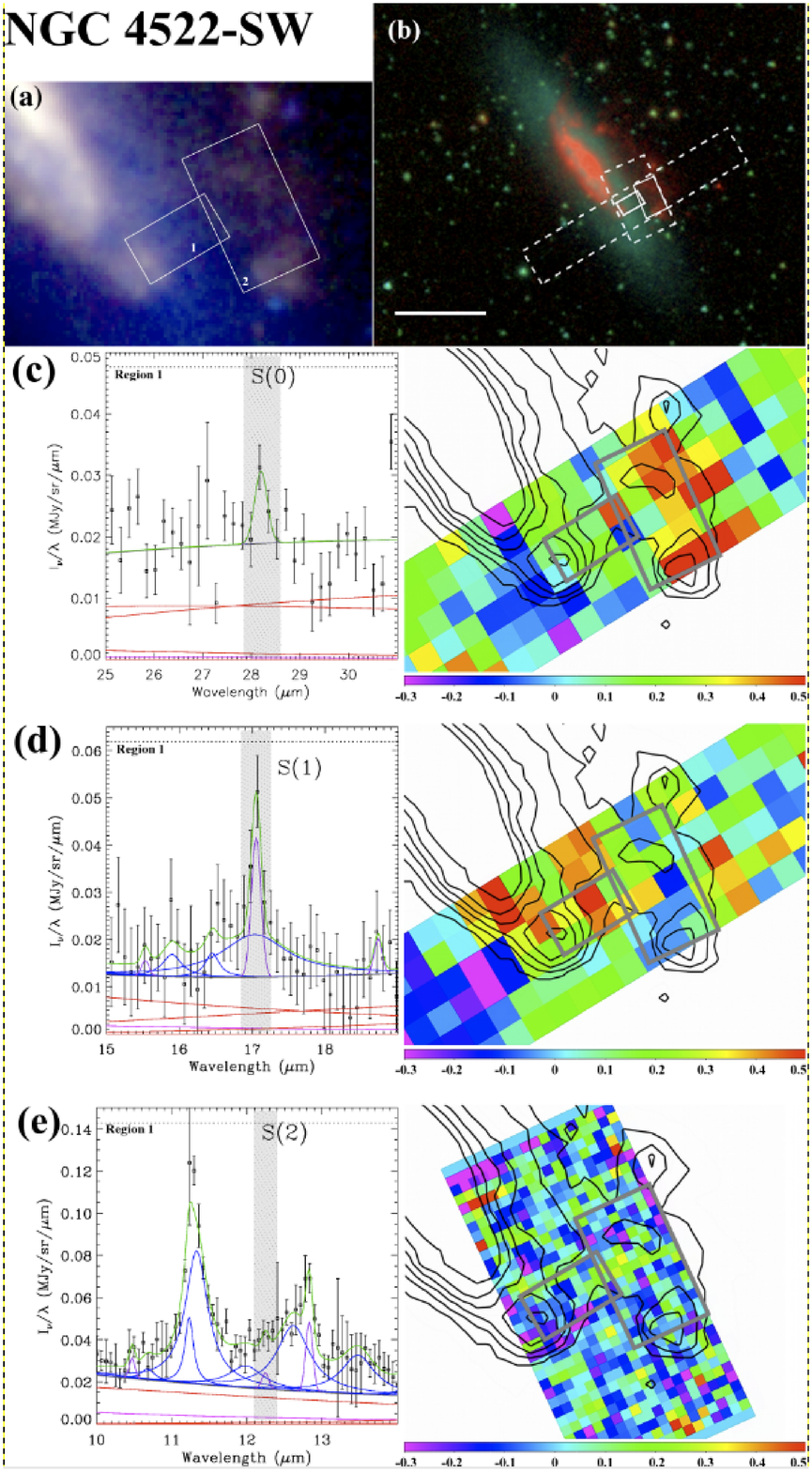}\\
\end{tabular}
\end{center}
\begin{center}
Figure~\ref{map_panel} (continued... NGC 4522 - southwest region). Note that the spectra in panels (c), (d) and (e) are from Region 1 (east of Region 2).
\end{center}
\vspace{0.5cm}
\end{figure*}

\section{Spectral analysis and discussion}
%  The first subsection probes the temperature of the observed rotational 
%\htwo\ emission via excitation diagrams.  We examine the possibility of shocks as the heating mechanism for 
%the observed \htwo\ transition emission lines by studying the ratio between the low energy \htwo\ transitions 
%($S0$, $S1$ and $S2$) and the PAH emission.
%The temperature distribution estimations of the warm molecular gas phase ($T$$\sim$100--1000 K)
%can be made from observations of rotational transitions of molecular hydrogen since they account for a significant volume
%fraction of the molecular clouds \citep{roussel07}.   
 Does the ram pressure interaction produce shocks which are strong enough 
to significantly enhance the \htwo/PAH ratios above the typical values expected from PDRs of star-forming
galaxies in the field? Section 4.1 examines the warm \htwo/PAH ratios of our sampled regions and compares them to 
those of a representative sample of field galaxies from the SINGS survey.  We also investigate the \Fe/\Ne\ and \Si/\Sulphur\ 
 emission line ratios for possible shock enhancements in Section 4.2.

Spectral mapping of the warm \htwo\ emission (Section 3.2) hint at N4522\_SW2 having lower \htwo\ temperatures 
relative to the other regions sampled  closer to the galactic disks.  In Section 4.3, we estimate the warm \htwo\ 
temperatures ($T= 100-1000$ K) by fitting simple one-temperature and  two-temperature
 models to the \htwo\ excitation diagrams derived from the observed \htwo\ rotational emission.

%In addition to FUV radiation from massive stars, shocks (originating from molecular outflows,
% supernova remnants or cloud collisions possibly due to a disturbed gravitational potential) may be responsible for
% heating the \htwo\ molecules into an excited state.  \citet{falgarone05} also showed that a small fraction of \htwo
% may be heated intermittently by supersonic turbulence pervading the entire ISM. In this section, we analyze the 
%rotational \htwo\  emission lines using excitation diagrams to estimate temperatures and densities of the warm and hot
% \htwo\ components. We also examine the possibility of shocks as the heating mechanism for the observed \htwo\ transition
% emission lines by studying the ratio between the low energy \htwo\ transitions ($S0$, $S1$ and $S2$) and the PAH emission.
\subsection{Ratio of warm \htwo\ to PAH emission}

If both the PAH and the warm \htwo\ emission originate from PDRs near where star formation is occuring, the 
ratio of warm \htwo\ to PAH should be fairly constant.  Indeed, recent observations of galaxies in the Local Universe, 
starburst galaxies and Seyfert galaxies have confirmed the strong correlation between the 8 $\mu$m PAH emission and 
the emission from \htwo\ rotational emission \citep[e.g.\ ][]{rigopoulou02,roussel07}.  If a fraction of the warm 
\htwo\ emission is heated intermittently
 by the dissipation of interstellar turbulence \citep{falgarone05} in addition to the excitation of cold 
\htwo\ via the UV emission of the young stellar population \citep[e.g.\ ][]{giard94}, such processes will
 be common in the ISM of both cluster and non-cluster galaxies. Therefore, the \htwo/PAH fractions should 
still be relatively similar between the cluster and non-cluster galaxies even if the intermittent interstellar
 turbulence introduces more scatter into the observed ratios.

%If the ram pressure interaction induces an increase in interstellar turbulence and 
%consequently, an elevated \htwo/PAH fraction; the root cause of the \htwo/PAH enhancement is still
% the process of ram pressure stripping.

Additional \htwo\ heating mechanisms such as shocks and gas heating via X-ray emission (mostly from AGNs) 
can produce significantly enhanced \htwo/PAH fractions relative to the typical fractions found in PDRs alone. 
The intense radiation fields which exist in the ionized regions around AGN \citep{desert88,voit92} are likely 
to excite the cold \htwo\ molecules to fluoresce and to radiatively cascade through the ro-vibrational 
levels of the ground electronic state \citep{roussel07}. Also, the AGN or supernova-induced shocks are more 
likely to destroy the PAH molecules while exciting more cold \htwo\ molecules into the warm phase \citep{roussel07}. 
 
As our regions do not contain any AGN nor strong X-ray emission, any enhancements in the ratio between
warm \htwo\ and PAH emission suggest the likely presence of shocks possibly due to ram pressure stripping.
Our measurements of the warm \htwo($S0$+$S1$+$S2$) and the 8 $\mu$m PAH emission for our sample are listed in 
Table~\ref{h2exc}. 
A range is provided for regions where significant detections were not measured in all 
three \htwo\ transitional states.  The lower bound of the range consists of the sum of the true detected 
values and the uncertainties of these measurements added in quadrature. The upper bound is the sum of the detected values 
and the 3$\sigma$ upper limits for non-detections.

\begin{figure}
\begin{center}
\includegraphics[scale=0.5]{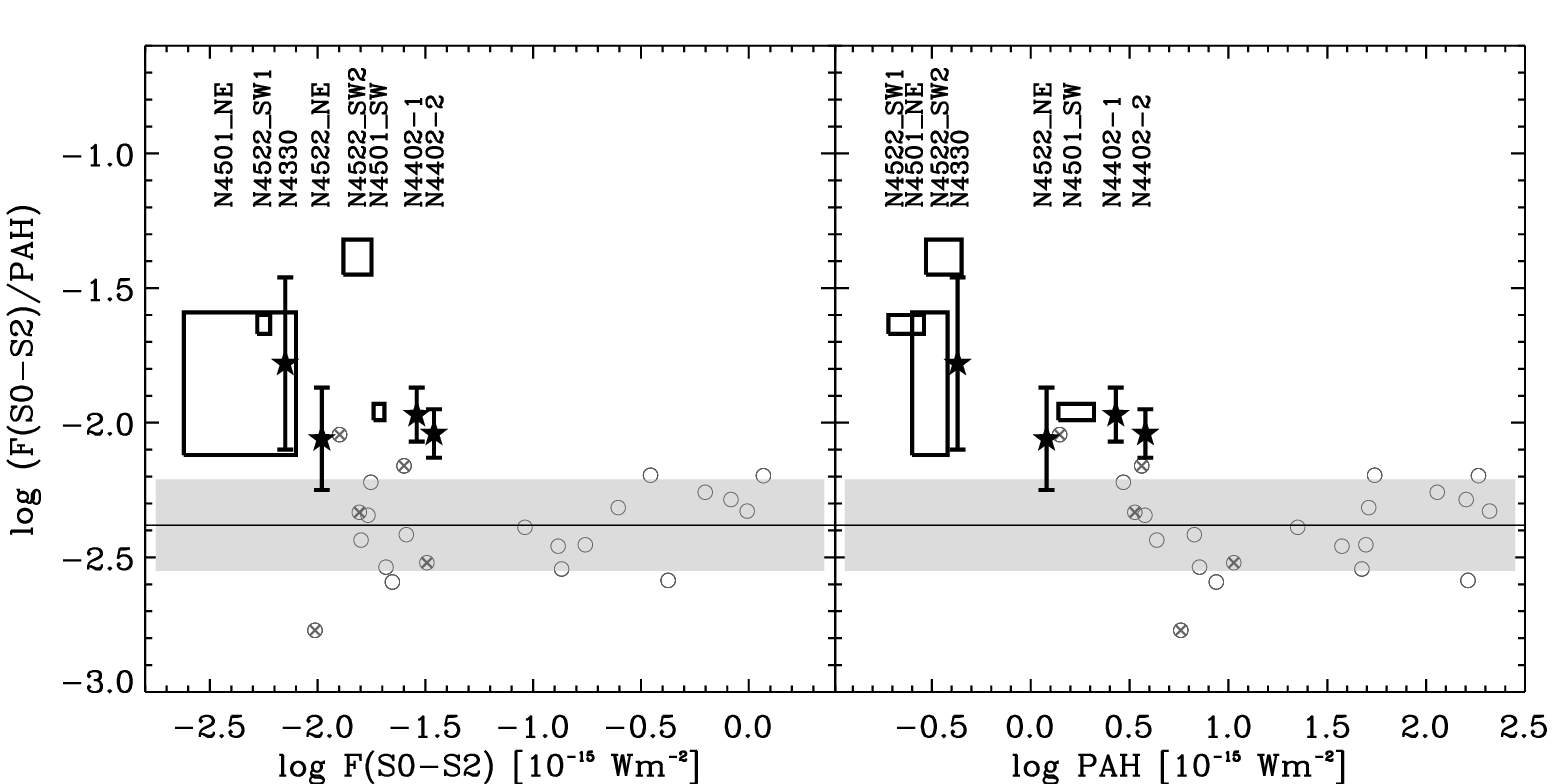}
\end{center}
\caption{The \htwo\ to PAH ratios as a function of \htwo\ (left) and PAH (right) of our observed regions are represented by 
the black star symbols and rectangles where the sides
of the rectangles mark the upper and lower limits.  Each of our observed regions can be identified by
the labels above the black symbols. The open circles and the $\otimes$ symbols represent the nuclear and non-nuclear regions
of the regions observed by the SINGS survey \citep{roussel07}, respectively. The gray shaded region shows the standard
deviation of the observed \htwo/PAH distribution around the mean (solid horizontal line)  for the SINGS sample.}
\label{h2pah}
\vspace{1cm}
\end{figure}

We compare our observations to those from the SINGS survey \citep{roussel07} in
Figure~\ref{h2pah}.  Our observations are represented by black star symbols, while the upper and lower limits of our
non-detections are represented by the open rectangles.
The publicly-available IRS spectra of the SINGS star-forming galaxies are processed using
the same method as for our sample (see Section 2.3).  The SINGS sample is divided into two types: nuclear (open circle)
and non-nuclear ($\otimes$ symbol) targets. The SINGS non-nuclear targets consist of extranuclear star-forming regions and the
nuclear targets often include AGN.  Our target apertures are comparable in size to those of SINGS which range 
in diameter from 60 pc to 3.8 kpc (with a median of 900 pc) depending on the target distances \citep{roussel07}.
 Note that only the sources from SINGS with significant detections of all three \htwo\ lower level transition lines are plotted. 

The \htwo/PAH ratios of the SINGS non-nuclear star-forming regions are very uniform and we find a mean 
log \htwo/PAH = $-2.37 (\pm 0.04)$ dex where the standard error of the mean represents the uncertainty in this 
average ratio.  On average, our observed regions have \htwo/PAH fractions  that are 2.6 times greater
(at a $\sim 5\sigma$ significance using the standard errors on the mean measurements) than 
those of the SINGS star-forming sample (see Figure~\ref{h2pah}).   The mean log \htwo/PAH fraction for 
our non-nuclear regions of galaxies experiencing ram pressure stripping  is $-1.96 (\pm 0.06)$ dex. 
%For the four detections in our sample, the standard deviation is log \htwo/PAH = 0.13 dex.

\begin{figure}
\begin{center}
\includegraphics[scale=0.5]{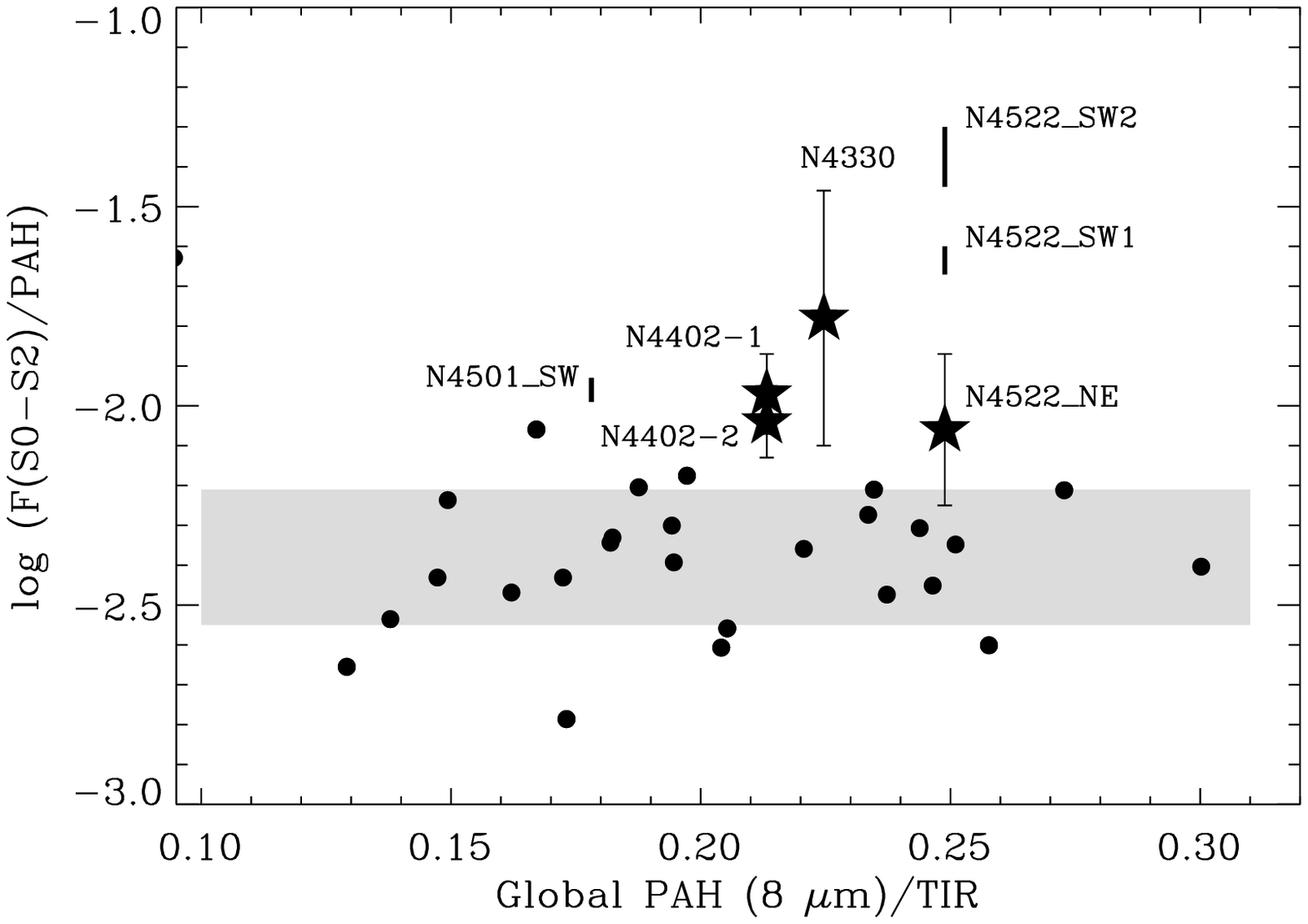}
\end{center}
\caption{The \htwo\ to PAH ratios as a function of  the global
PAH (8 $\mu$m)/TIR for our target regions as star symbols.  The measured PAH (8 $\mu$m)/TIR
for the SINGS targets which are neither Seyfert nor LINER targets are plotted as
solid black dots. The gray shaded region shows the standard
deviation of the observed \htwo/PAH distribution around the mean \htwo/PAH  for the SINGS sample. }
\label{bendo}
\vspace{1cm}
\end{figure}

Although it is possible that the elevated \htwo/PAH fractions from our sample is due to a PAH deficit,
 we show in Figure~\ref{bendo} that this is not the case. Previous studies \citep[e.g.\ ][]{haas02,bendo08} 
have shown that the PAH emission correlates well with the cold dust emission or the bolometric total 
3--1100 $\mu$m infrared emission (TIR) and that the the global PAH/TIR fraction is a good proxy for 
the intensity of the radiation field heating the diffuse ISM \citep{tielens99,helou01}. 
A comparison of the PAH/TIR ratios for our sample and those of the SINGS galaxies \citep{dale07} reveal that
our targeted regions are not deficient in PAH emission and in many cases are higher than those from the 
SINGS survey (see Figure~\ref{bendo}).  Hence, it is likely that the elevated \htwo/PAH fractions measured
in most of our sample is due to an excess of warm \htwo\ and not a PAH deficit.

It is interesting to note that we do not observe any significant differences in \htwo/PAH excess between the leading or 
trailing sides of the galaxy disks within our sample with the one exception being the  extraplanar 
 star-forming region in NGC 4522 (N4522\_SW2).  We hypothesize that the very high \htwo/PAH fraction is probably 
due to both an enhancement in warm \htwo\ as well as a PAH deficit because the $S(0)/S(1)$ fraction found for this
region (see Section 3.2) suggests that the  density of star forming regions is much lower in the extraplanar 
region than the disk regions.  As such, we expect that this will result in weaker PAH emission.

%The intense radiation fields which exist in the cores of \HII\ regions \citep{giard94} and ionized regions 
% In Section 4.2, we find enhanced ratios of \Fe/\Ne\ for our sample even though the \Si/\Sulphur\ ratios
% are consistent with those found in star-forming regions rather than AGN. As ram pressure stripping 
%is the dominant physical process affecting our entire sample, we think that {\em{it is likely that the enhanced
%\htwo/PAH ratios (relative to star-forming galaxies from SINGS) are a result of additional shock heating 
%from the ram pressure interaction}}.  See Section 4.2 for further details of enhanced \Fe/\Ne\ ratios in 
%our sample.

Compared to the ram pressure-driven shocks observed in the Norma cluster galaxy, ESO 137-001 
\citep{sivanandam09},  our Virgo sample show a more modest excess in shock-excited \htwo. In the extraplanar
star-forming tail of ESO 137-001, \citet{sivanandam09} found an excess of warm \htwo\ at an equal or 
greater level to that found in AGN regions within the SINGS sample. This result is sensible since the Coma and Norma clusters have much higher masses and generally stronger ram pressures than the Virgo cluster  \citep{sun07,tonnesen12}.
Follow-up observations which probe to fainter surface brightness limits in warm \htwo\ over a larger
field-of-view will provide further verification of our observed excess in the \htwo/PAH fractions as
the observed ratios are very sensitive to the presence or absence of any diffuse emission. 
 Alternatively, observations of other shock indicators will provide independent tests for the 
existence of ram pressure-induced shocks in the ISM of in-falling galaxies.

%the warm \htwo\ excess.   Deeper observations of PAH will also verify the existence of any PAH deficit
%Since our results are hampered by the low SNR nature of our observations and are only suggestive of ram pressure-induced
%shocks,  follow-up observations which probe to lower surface brightnesses in warm \htwo\ will provide verification of 
%the warm \htwo\ excess.   Deeper observations of PAH will also verify the existence of any PAH deficits.
% Alternatively, further observations of other shock indicators such as the ratio between $[$C {\sc{ii}}$]_{158\mu m}$ and 
%$[$O {\sc{i}}$]_{63\mu m}$ will provide independent verification of the existence of shocks near the leading
%edges of ram pressure stripping. We have recently submitted a proposal aimed at measuring these two emission lines (in 
%N4402, N4501\_SW \& N4522\_NE) using the PACS spectrometer on board the Herschel Space Telescope to explore this
%possibility.

\begin{table}
\scriptsize{
\begin{center}
\caption{\htwo($S0+S1+S2$)/PAH ratios of our observed sample.}
\label{h2exc}
\begin{tabular}{lcc}
\tableline
\tableline
 Object & log (F($S0+S1+S2$))& log (F($S0+S1+S2$)/PAH) \\
 & ($\times 10^{-15}$ W m$^{-2}$) & \\
\tableline
N4330 & $-2.15 \pm 0.30$ & $-1.78 \pm 0.32$\\
N4402\_1 & $-1.54 \pm 0.07$ & $-1.97 \pm 0.10$\\
N4402\_2 & $-1.46 \pm 0.04$ & $-2.04 \pm 0.09$\\
N4501\_SW & $-1.86$, $-1.69$ & $-2.10$,  $-1.93$\\
N4501\_NE & $-2.72$,  $-2.10$& $-2.22$, $-1.59$\\
N4522\_NE &$-1.98 \pm 0.14$ & $-2.06 \pm 0.19$\\
N4522\_SW1&$-2.36$, $-2.22$ & $-1.74$, $-1.60$\\
N4522\_SW2&$-1.88$, $-1.75$ & $-1.45$, $-1.32$\\
\tableline 
\end{tabular}
\end{center}}
\end{table}
\vspace{0.5cm}

\subsection{Emission line ratios of \Fe/\Ne\ and \Si/\Sulphur}

\begin{figure}
\begin{center}
\includegraphics[scale=0.8]{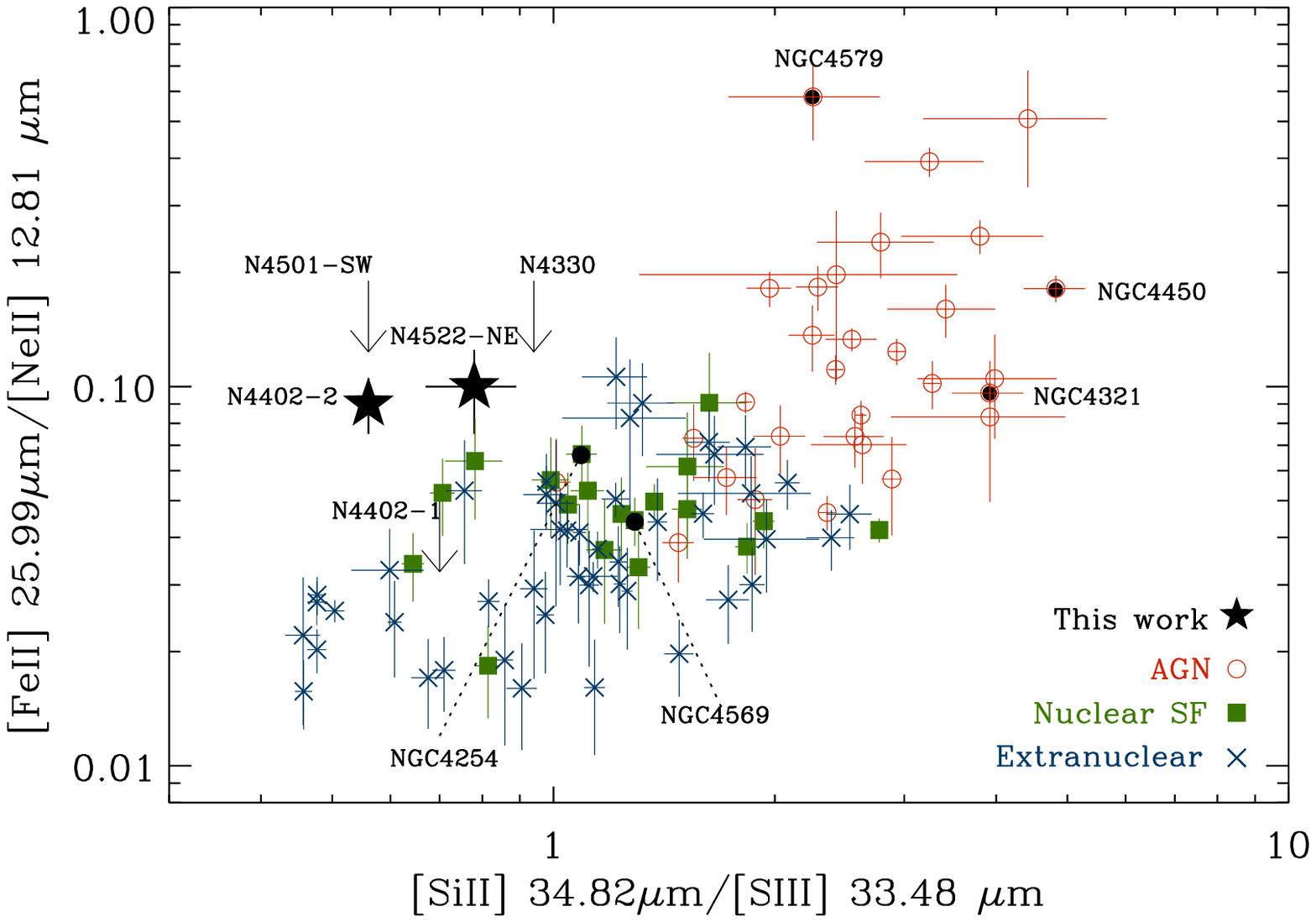}\\
\end{center}
\caption{Emission line ratios \Fe/\Ne\ versus \Si/\Sulphur\ for our sample (black star
symbols and black upper limits). The measured line ratios from the SINGS survey of AGN, nuclear star-forming regions 
and extranuclear star-forming regions are represented by the orange open circles, green solid squares
and the blue crosses, respectively  \citep{dale09}. The black solid circles represent SINGS galaxies that are in
the Virgo Cluster. }
\label{fenesis}
\end{figure}

In this section, we examine the  ratios of the MIR fine structure atomic emission lines
\Fe/\Ne\ and \Si/\Sulphur\, for possible evidence of shock-heating in the ISM.
  Typically, \Ne\ and \Sulphur\ have ionization potentials of approximately 22--23 eV 
and originate from within \HII\ regions; while \Fe\ and \Si\ have lower ionization potentials ($\sim$8 eV) 
and are emitted from outside \HII\ regions.   Figure~\ref{fenesis} plots \Fe/\Ne\ as a function of \Si/\Sulphur\ 
for our sample (represented by star symbols and upper limits) as well as the observed ratios for a sample of 
AGN and star-forming regions from the SINGS survey \citep{dale09}.  The emission line ratios are distinctly 
different between AGNs and star-forming regions---AGNs have higher line ratios.  This difference is not well 
understood in detail, but is believed to be the result of either: (i) the enhanced liberation of dust grain 
constituents in AGN regions producing stronger Fe and Si lines; (ii) X-rays from AGN producing large volumes of low 
ionization gas (including enhanced $[$\Fe$]$ 34.82 $\mu$m and $[$\Si$]$ 25.99 $\mu$m emission); or (iii) higher 
gas densities in AGNs than star forming regions \citep{dale09}.

In our sampled regions, we were able to detect these emission lines in the disk of NGC 4402 (N4402\_2)
and on the leading side of the disk of NGC 4522 (N4522\_NE). Our \Si/\Sulphur\ ratios  are consistent
with star-forming regions from the SINGS sample but our \Fe/\Ne\ ratios are somewhat enhanced
relative to the SINGS star-forming regions. While metallicity affects these line ratios \citep{ohalloran06,dale09},
 the metallicities of our Virgo sample are between 0.5--1 $Z_{\odot}$ \citep{crowl08}---within the range of
metallicities of the extranuclear star-forming regions in SINGS \citep{dale09}.  As our galaxies are within 
100--200 Myr from peak ram pressure \citep{crowl08}, it is possible that the increased  \Fe/\Ne\ ratios 
relative to \Si/\Sulphur\ are the result of interaction-driven shocks returning \Fe\ and \Si\ from the 
dust grains back to the ISM.  In the ISM of the Milky Way, there is a strong depletion of Fe onto grains 
\citep[e.g.\ ][]{savage96}, and gas phase Fe is largely due to the processing of grains by shocks 
\citep[e.g.\ ][]{alonso97,alonso03,forbes93,ohalloran06,armus07}.
Since Fe is approximately 7 times more easily depleted than Si in the Milky Way \citep{draine04}, a shock 
passing through the dusty ISM will release $\sim 7$ times more Fe than Si, possibly driving the ratio in 
the observed direction. If so, this suggests that the difference in line ratios between AGN and star-forming 
regions is not due to the enhanced liberation of atoms from dust grains in AGN, but is due to one of the other 
possibilities.

As can be seen in Figure~\ref{fenesis}, the SINGS sample includes several other Virgo cluster galaxies not in
our sample.  These other Virgo galaxies are not currently experiencing strong ram pressure stripping and lie within 
the core distribution of galaxies for both nuclear and extranuclear regions alike.  As such, we posit that the 
elevated \Fe/\Ne\ ratios observed in our sample is due to an enhanced depletion of Fe, probably from shocks driven
by ongoing ram pressure since ram pressure is the dominant physical process affecting our entire sample.

\subsection{Excitation diagrams}

\begin{figure*}
\begin{center}
\includegraphics[scale=0.55]{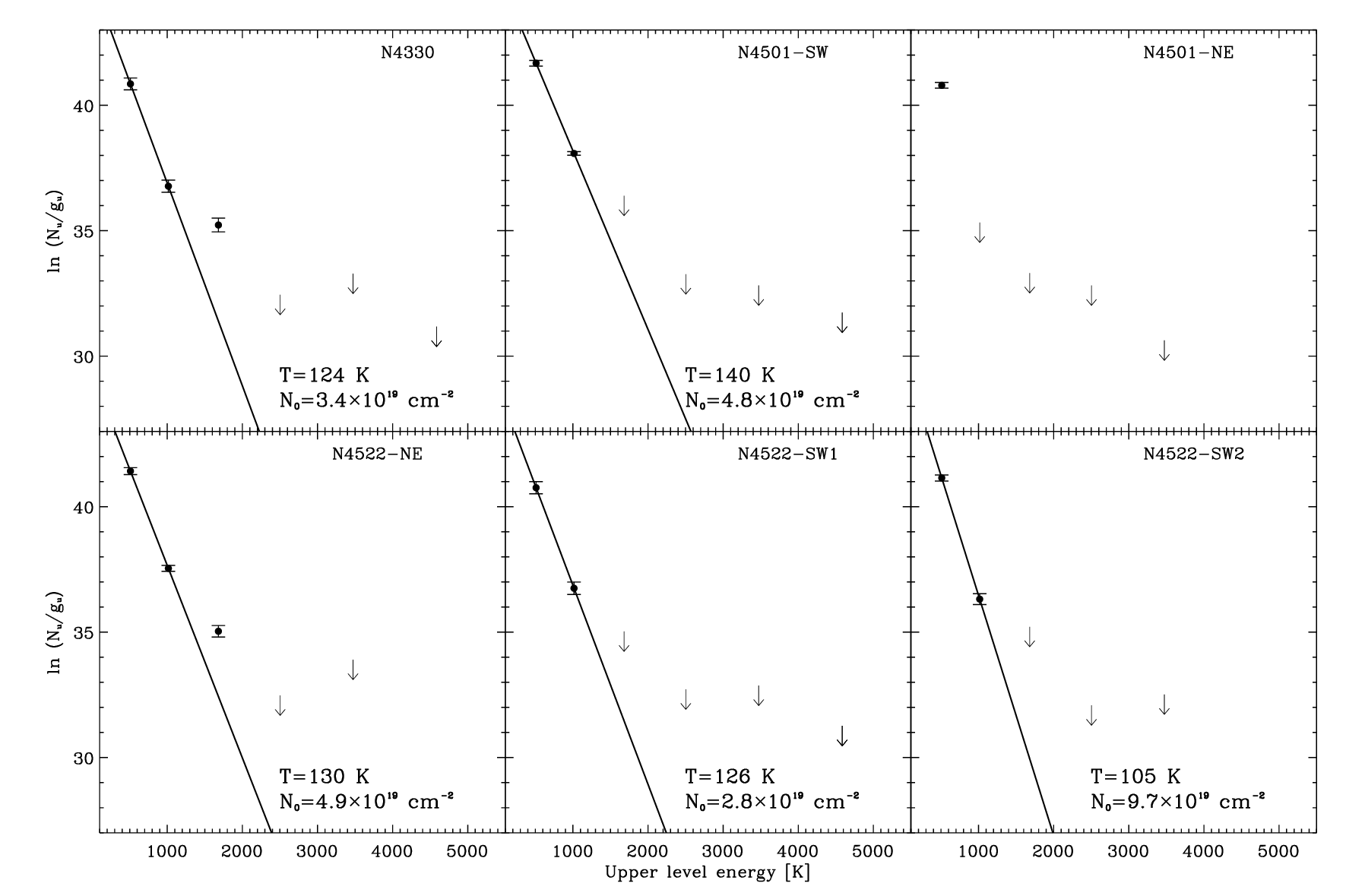}
\end{center}
\caption{Excitation diagrams of our Virgo galaxy sample (N4522\_NE) known to be experiencing ram pressure stripping
 where thermalization can be assumed and where $\rm{OPR}=3$.  Two-temperature models are not estimated for N4330,
N4501\_SW, N4501\_NE, N4522\_SW1 and N4522\_SW2 as we only obtained upper limits for the lowest transition states.  Therefore
the single-temperature fits are only approximations to the true temperatures in these targets.
 }
\label{ltediag}
\vspace{1.1cm}
\end{figure*}

%the observed \htwo\ transition emission lines by studying the ratio between the low energy \htwo\ transitions 
%($S0$, $S1$ and $S2$) and the PAH emission
%This section analyzes the observed emission line intensities of our sample and specifically explores the idea that the 
%heating mechanism of the warm \htwo\ emission observed near the leading edges of ram pressure interaction may be 
%due to ram pressure-induced shocks. We begin by analyzing the emission from the rotational \htwo\ lines using 
%excitation diagrams to constrain the temperatures and densities of the warm and hot \htwo\ components. 
Do the elevated ratios of \htwo/PAH and \Fe/\Ne\ suggest warmer ISM conditions relative
to other star-forming galaxies? Are the ISM conditions different between the leading and trailing 
sides of a galaxy experiencing ram pressure stripping?
If local thermodynamic equilibrium (LTE) is valid for our regions, we can constrain the temperatures and
densities of the warm and hot \htwo\ components by fitting simple one- and two-temperature models to
to the excitation diagrams.
Excitation diagrams (see Figure~\ref{ltediag}) illustrate the distribution of the level populations 
by mapping $N_u/g_u$ (where $N_u$ is the molecular column density and $g_u$ is the statistical weight 
for that transition) as a function of the upper level energy ($E_u$). 

Previous observations of starburst galaxies \citep[e.g.\ ][]{rigopoulou02} and ultraluminous galaxies 
\citep[e.g.\ ][]{higdon06} deduced that the lowest rotational levels are in LTE   because the warm 
\htwo\ emission is likely to originate mainly from the densest PDRs with densities greater than 
$10^3$ cm$^{-3}$ \citep{burton92,kaufman06}.  However, recent observations of nearby galaxies in the
 SINGS sample found that the rotational transitions may not behave monotonically as a function of energy 
level---indicative of a departure from thermalization \citep{roussel07}. Using the method outlined by 
\citet{roussel07}, the apparent transition temperatures should satisfy the following conditions if the
 observed gas is indeed at LTE and have ortho-to-para ratios (OPR) of 3:  
$T(S0-S1) \leq T(S0-S2) \leq T(S1-S2) \leq T(S1-S3) \leq T(S2-S3)$. However it should be noted that it is
still possible to satisfy this condition for slight departures from local thermalization.

The excitation temperatures can be determined for each pair of transitions via: 
\begin{equation}
kT_{\mathrm{ex}} = (E_{u2}-E_{u1})/ {\mathrm{ln}}(N_{u1}/N_{u2} \times g_{u2}/g_{u1})
\end{equation} where $N_{u1}/N_{u2} = F_1/F_2 \times A_2/A_1 \times \lambda_1/\lambda_2$ \citep{roussel07}.  By 
assuming that the \htwo\ gas is in local thermodynamic equilibrium (LTE), the total column density, $N_{\rm{T}}$ can be 
derived from $N_u=g_uN_{\rm{T}}\, {\rm{exp}}[-E_u/(kT)]/Z(t)$, where $Z(T) \approx 0.0247T/[1- {\rm{exp}}(-6000 {\rm{K}}/T)]$ 
is the partition function for temperatures greater than 40 K. In particular, the condition 
$T_{\mathrm{ex}}(S1-S2) <T_{\mathrm{ex}}(S1-S3) < T_{\mathrm{ex}}(S2-S3)$ has to be satisfied if thermalization is attained.  It 
should be noted that $T_{\mathrm{ex}}(S0-S2)$ and  $T_{\mathrm{ex}}(S1-S3)$ are determined directly from the observed emission 
line fluxes and hence, independent of the OPRs at the high-temperature limit.
 
\begin{table*}
\begin{center}
\caption{Apparent \htwo\ transition temperatures of our sample.}
\vspace{0.2cm}
\label{ltetest}
%\begin{tabular}{lccccc}
%\tableline
%\tableline
%Target & $T(S0-S1)$ & $T(S0-S2)$ & $T(S1-S2)$ & $T(S1-S3)$ & $T(S2-S3)$ \\
% & (K) &(K) & (K) & (K) & (K) \\
%\tableline
%N4330 & 104 $\pm11$& 208 $\pm15$& 369 $\pm83$& --- &---\\
%N4402\_1 & 120 $\pm9$&209 $\pm11$& 274 $\pm52$& 276$ \pm2$ & 245$\pm11$\\
%N4402\_2 & 122 $\pm8$&210 $\pm4$& 268 $\pm18$& 296$ \pm2$ & 279$\pm8$\\
%N4501\_SW &116 $\pm9$&--- & --- & --- &---\\
%N4501\_NE & --- & --- & --- & --- & ---\\
%N4522\_NE & 109 $\pm8$& 183 $\pm5$& 241 $\pm56$& --- &--- \\
%N4522\_SW & 106 $\pm12$& --- & --- & --- &---\\
\begin{tabular}{lccccc}
\tableline
\tableline
Target & $T(S0-S1)$ & $T(S0-S2)$ & $T(S1-S2)$& $T(S1-S3)$ & $T(S2-S3)$  \\
 & (K) &(K) & (K) &(K) & (K) \\
\tableline
\multicolumn{6}{c}{\em{OPR = 3.0}} \\
N4330 & 104 $\pm11$& 208 $\pm15$& 369 $\pm83$ & $<$344  &$<$282 \\
N4402\_1 & 120 $\pm9$&209 $\pm11$& 274 $\pm52$ &276 $\pm2$ &245 $\pm11$\\
N4402\_2 & 122 $\pm8$&210 $\pm4$& 268 $\pm18$ & 296 $\pm2$ & 279 $\pm8$\\
N4501\_SW &116 $\pm9$&$<$222 & $<$342 &$<$309 & --- \\
N4501\_NE & $<$99 & $<$214 & ---&--- & ---\\
N4522\_NE & 109 $\pm8$& 183 $\pm5$& 241 $\pm56$ &$<$294 &$<$305\\
N4522\_SW1 & 106 $\pm12$& $<$204&$<$336&$<$369 & --- \\
N4522\_SW2 & 90 $\pm3$& $<$197&$<$488&$<$352 & --- \\
\tableline
\multicolumn{6}{c}{\em{OPR = 1.7}} \\
N4330 & 118 $\pm12$& 208 $\pm15$& 280 $\pm63$ & $<$344  &$<$350 \\
\tableline
\multicolumn{6}{c}{\em{OPR = 2.0}} \\
N4402\_1 & 133 $\pm10$&209 $\pm11$& 234 $\pm44$ &276 $\pm2$ &279 $\pm13$\\
\tableline
\multicolumn{6}{c}{\em{OPR = 2.5}} \\
N4402\_2 & 128 $\pm8$&210 $\pm4$& 250 $\pm16$ & 296 $\pm2$ & 297 $\pm8$\\
\tableline
\tableline
\end{tabular}
\end{center}
\end{table*}

%have apparent temperatures of consecutive rotational transitions which 
%are not monotonic as a function of upper level energy and hence, a departure from thermalization between the ortho and 
%para levels where the equilibrium ratio is 3 \citep{roussel07}.  
%An ortho-to-para (OPR) ratio in the excited states which differ from the equilibrium value 
%(OPR = 3) can indicate a deviation from thermalization.

\begin{figure*}
\begin{center}
\includegraphics[scale=0.5]{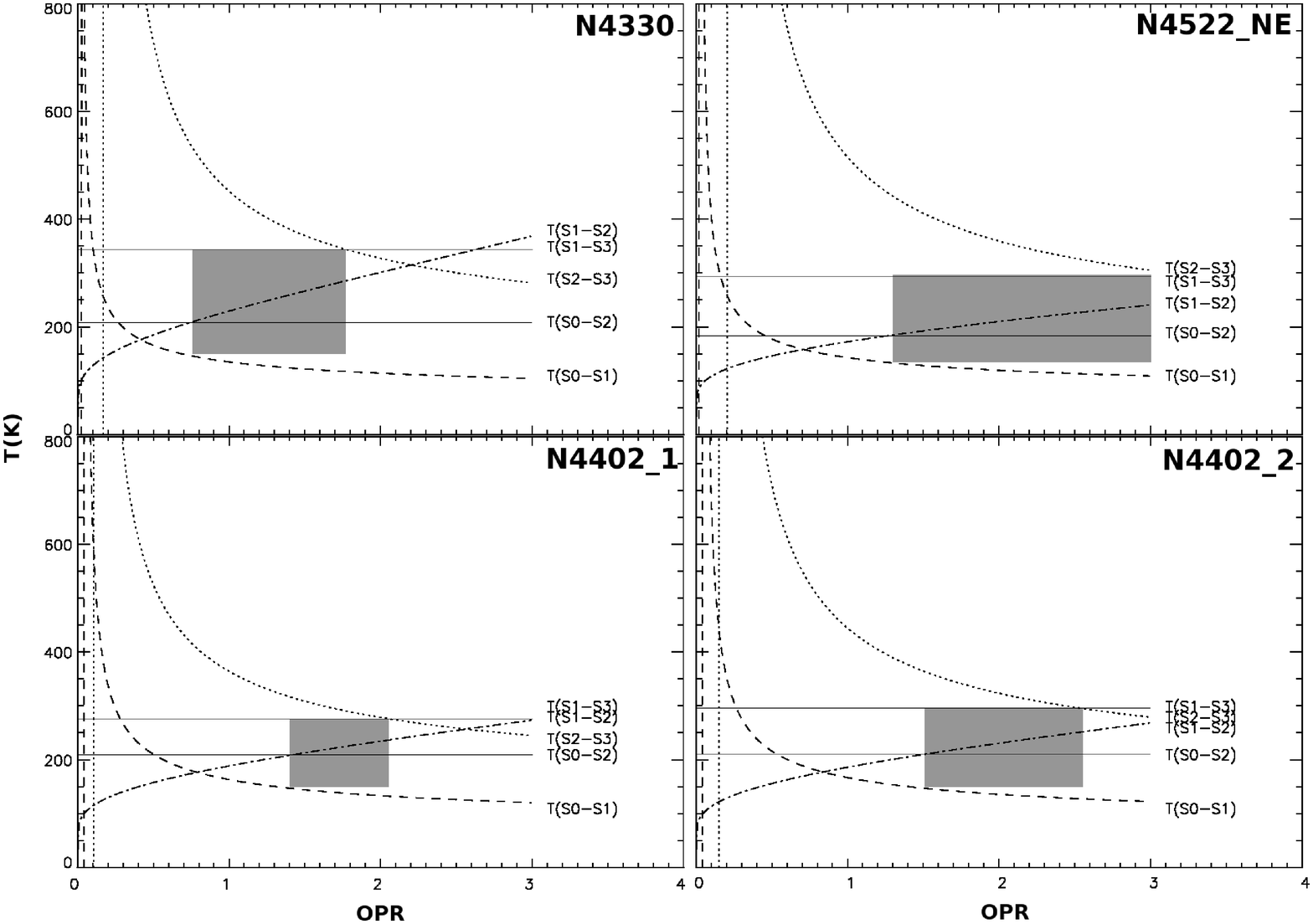}
\end{center}
\caption{Apparent temperatures as a function of the ortho-to-para ratios (OPR).  The possible range of OPR for 
each of our targets are indicated by the gray shaded regions. }
\label{opr}
\vspace{1.1cm}
\end{figure*}

Table~\ref{ltetest} lists the estimated apparent excitation temperatures for our sample and the shaded regions 
in Figure~\ref{opr} show  the possible range of OPR values for these targets.  It should be noted that the 
the gray shaded regions  do not include the uncertainties in our measurements.  The conditions proposed 
by \citet{roussel07} for thermalization with an OPR of 3 can be  satisfied by N4330 and N4522\_NE when the 
uncertainties are taken into consideration. However in N4402\_1 and N4402\_2, the  apparent temperatures of 
consecutive rotational transitions do not appear to behave monotonically as a function of upper level energy
 and suggest OPR values which are less than 3.

% The measurements for N4330, N4402\_1, N4402\_2 and N4522\_NE  only appear to satisfy the proposed conditions 
%for LTE if the $S(3)$ measurements are omitted (due to the high uncertainty in the $S(3)$ measurement).
%However, it should be noted that the degeneracy between mass and temperature can only be lifted if good detections of
% higher-order emission (e.g.\ $S(3)$) are available.  

%On the other hand, the estimated temperatures for the $S(3)$ higher order transition  (see Table~\ref{ltetest}) are
% fairly uncertain.  Therefore, 
Assuming local thermalization, we can approximate a single temperature that corresponds to the lower
energy transitions at $S(0)$ and $S(1)$ of our targets via 
${\rm{ln}} (N(T_u)/g_u) = {\mathrm{ln}}(N_{0,\rm{ONE}}) - \frac{T_u}{T_{\rm{ONE}}}$ 
where $N_{0,{\rm{ONE}}}$ is the column density of the $J=0$ state, $T_u$ is the upper energy level of the transition 
and $T_{\rm{ONE}}$ is the temperature of the warm \htwo\ in the $J=0$ state \citep{roussel07}. Table~\ref{fittemp} 
lists the derived temperatures and column densities from these single-temperature fits for our observed regions.
Bootstrap resampling is used to estimate the uncertainties in the derived properties. 

In reality, the observed emission is likely to arise from gas at a range of temperatures. Previous studies \citep[e.g.\ ][]{roussel07, appleton06}
 found a two-temperature model to be sufficient for characterizing the observed excitation diagrams based on a limited number of lines. 
%Therefore, where local thermalization (OPR=3) can be assumed, we can perform a Levenberg-Marquardt least squares fit to the following
%two-temperature model:
%\begin{eqnarray}
%{\mathrm{ln}}(N(T_u)/g_u &=& {\mathrm{ln}} \Big[{\mathrm{exp}} \Big(-\frac{T_u}{T_1}\Big) + f_{2,1} \, {\mathrm{exp}} \Big(-\frac{T_u}{T_2}\Big)\Big] + {\mathrm{ln}}(N_{0,1})
%\end{eqnarray}
%where $T_1$ and $T_2$ are the temperatures of the first and second components; $N_{0,1}$ represents the column density 
%of the $J=0$ state of the first component; and $f_{2,1} = N_{0,2}/N_{0,1}$. 
However,  to solve for the four parameters in a two-temperature model, we need to detect at least all four lowest transition 
states ($S(0)$, $S(1)$, $S(2)$ and $S(3)$).  As we have not detected any $S(3)$ emission in our sample of regions where
thermalization can be assumed, we constrain the temperatures and column densities of the warm \htwo\ components via 
the simpler single-temperature model.   Figure~\ref{ltediag} shows the 
excitation diagrams for the observed \htwo\ rotational emission detected in N4330, N4501\_SW, N4501\_NE,
N4522\_NE, N4522\_SW1 and N4522\_SW2. The single-temperature fits of the warm \htwo\ components are represented by the
solid black line.  The temperature range of the warm component for the sample shown in Figure~\ref{ltediag} is approximately 105--140 K.  

We do not observe a significant difference in the warm \htwo\ temperatures on the leading versus the trailing sides of 
galaxies with the exception of the extraplanar star-forming region on the trailing side of NGC 4522 (N4522\_SW2). 
Consistent with our results in Section 3.2, we find that the warm \htwo\ in the extraplanar region of NGC 4522 is indeed
 $\sim$20\% colder than the other sampled regions within the disk of NGC 4522 (N4522\_NE and N4522\_SW1).
 On the other hand, the column density
of \htwo\ in the $J=0$ state for N4522\_SW2 is also a factor of 2 to 3 times greater than those of N4522\_NE and N4522\_SW1.
%In fact, the $N_{0,1}$ of N4522\_SW2 is the highest of all the regions sampled in this paper.

\begin{figure*}
\begin{center}
\includegraphics[scale=1.]{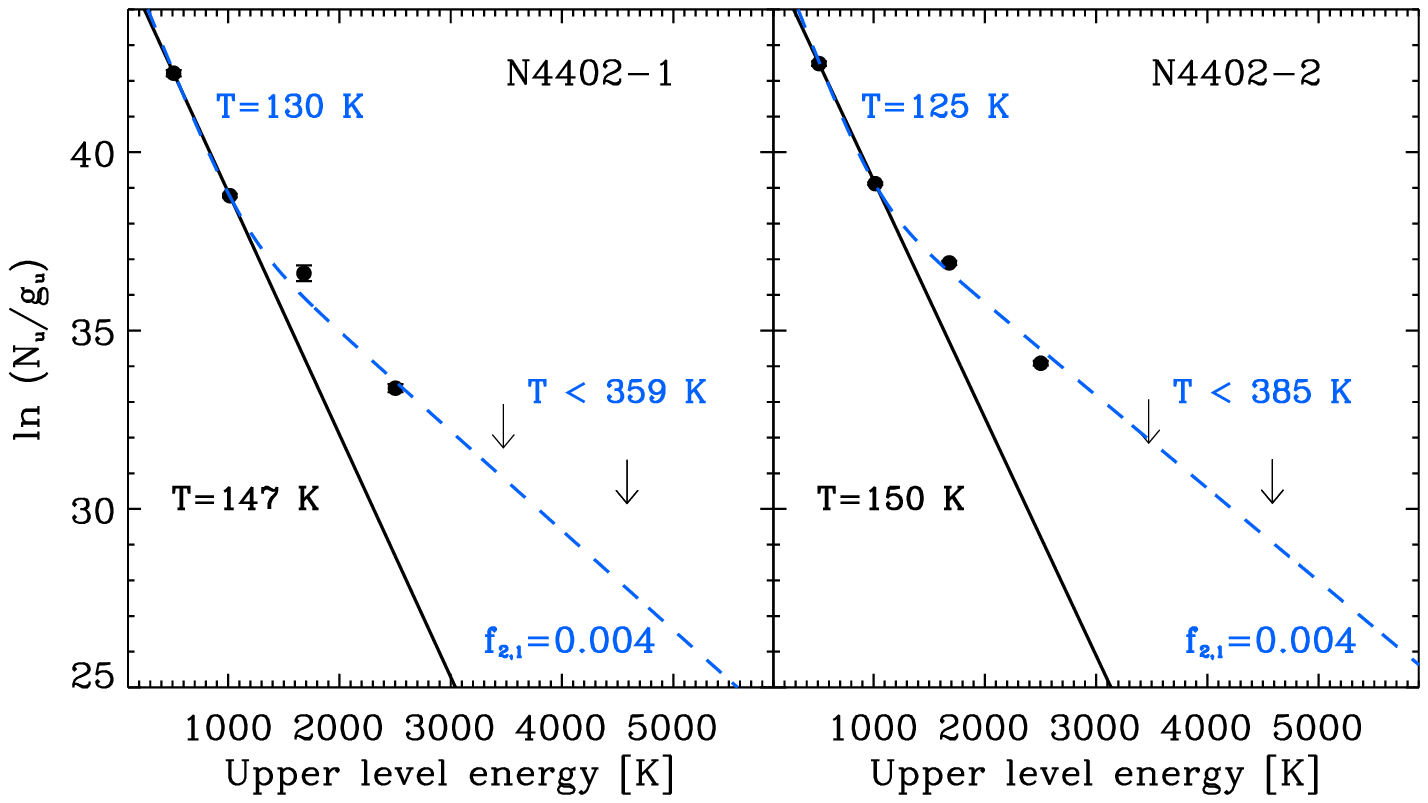}
\end{center}
\caption{Excitation diagrams of our Virgo galaxy sample (N4402\_1, N4402\_2) where the $\rm{OPR}$ is likely to be less than 3.  
 The two-temperature models (with the two temperatures, T$_1$ and T$_2$,  written in blue) are shown by blue dashed lines and the single 
temperature model (with the temperature printed in black) for the lower energy transitions are shown by the black solid line.   The column density fraction between 
the hot and warm phase of \htwo\ is given by $f(2-1)$. }
\label{ltediaglim}
\vspace{1.1cm}
\end{figure*}

To estimate the gas temperatures and column densities of gas from the observations of  N4402\_1 and N4402\_2, 
we adopt the method described by \citet{roussel07} where we fix the lower and upper temperatures in the 
two-temperature models to be at 0.98 $T(S0-S1)$ and 1.3 $T(S1-S3)$, respectively.  Using the OPR values described in
Table~\ref{ltetest} and the fixed temperatures to constrain the two-temperature fits, we estimate the 
hot-to-warm gas density ratio to be  0.004 for both the targeted regions in N4402, respectively.  
The results of these constrained fits can be found in Table~\ref{fittemp}. The energy diagrams of  
N4402\_1 and N4402\_2 overlaid with these constrained temperature models are presented in Figure~\ref{ltediaglim}.
As the two-temperature models allow for a small contribution from the hotter gas to the $S(0)$ and $S(1)$ emission,
 the warm components from the two-temperature models typically have lower temperatures and higher column densities
than those found from the the single-temperature models.

\begin{table*}
\footnotesize{
\begin{center}
\caption{Fitted temperatures for the warm \htwo\ observed in our sample.  }
\vspace{0.2cm}
\label{fittemp}
\begin{tabular}{lcccccc}%c}
\tableline
\tableline
Target & $T_{\rm{ONE}}$ & $N_{0,\rm{ONE}}$&$T_{\rm{TWO},1}$ & $T_{\rm{TWO},2}$ & $N_{0,1}$& $f_{2,1}$ \\%& $N_{\rm{TOT}}$  \\
 & (K) & $\times 10^{19}$ cm$^{-2}$ & (K) & (K) & $\times 10^{19}$ cm$^{-2}$&  \\%&$\times 10^{19}$ cm$^{-2}$\\
\multicolumn{1}{c}{(1)} & (2) & (3) &(4) &(5) &(6) & (7) \\%& (8) \\
\tableline
%N4330** & 124$^{+14}_{-13}$& 3.4$^{+3.4}_{-1.7}$& 62$^{+1}_{-4}$& 347$^{+9}_{-7}$ &180$^{+169}_{-33}$ &1.07$^{+0.07}_{-0.46} \times 10^{-4}$ \\%& 10.2$^{+24.0}_{-6.2}$\\
%N4402--1** & 147$^{+5}_{-6}$ & 6.9$^{+1.9}_{-1.4}$ & 65$^{+3}_{-34}$ & 278$^{+2}_{-41}$ &  308$^{+132}_{-305}$ &0.0009$^{+0.2141}_{-0.0003} $ \\%& 27.3$^{+15.0}_{-9.6}$ \\
%N4402--2** &150$^{+5}_{-5}$  & 8.4$^{+1.7}_{-1.4}$ & 66$^{+2}_{-35}$ & 297$^{+1}_{-42}$&502$^{+106}_{-496}$&0.0006$^{+0.1174}_{-0.0003}$\\%& ---\\
%N4330** & 124$^{+14}_{-13}$& 3.4$^{+3.4}_{-1.7}$& 116& $<447$ &3.8 &0.002 \\%& 10.2$^{+24.0}_{-6.2}$\\
N4330 & 124$^{+14}_{-13}$& 3.4$^{+3.4}_{-1.7}$&--& --- &---&---\\
N4402--1** & 147$^{+5}_{-6}$ & 6.9$^{+1.9}_{-1.4}$ & 130& $<359$ &  10.7 &0.004 \\%& 27.3$^{+15.0}_{-9.6}$ \\
N4402--2** &150$^{+5}_{-5}$  & 8.4$^{+1.7}_{-1.4}$ & 125 & $<385$ &16.4&0.004\\%& ---\\
N4501\_SW &140$^{+7}_{-6}$ & 4.8$^{+1.7}_{-1.3}$ &---&  ---& --- &---\\%&  16.8$^{+8.4}_{-5.1}$\\
N4501\_NE & ---& ---& ---& --- &---&---\\%& ---\\
%N4522\_NE & 130$^{+8}_{-8}$ & 4.9$^{+8.9}_{-0.5}$ & 105 $^{+2}_{-4}$ & 322 $^{+15}_{-14}$ &  1.2$^{+0.4}_{-0.3}$ &0.0025 $^{+0.0006}_{-0.0005}$\\%& 13.0$^{+9.2}_{-5.2}$  \\
N4522\_NE & 130$^{+8}_{-8}$ & 4.9$^{+8.9}_{-0.5}$ &  ---& --- &---&---\\
N4522\_SW1 &126$^{+16}_{-13}$ & 2.8$^{+2.9}_{-1.4}$ &  ---& --- &---&---\\%&  9.0$^{+22.8}_{-5.5}$\\
N4522\_SW2 &105$^{+6}_{-7}$ & 9.7$^{+5.0}_{-3.3}$ &  ---& --- &---&---\\%&  9.0$^{+22.8}_{-5.5}$\\
\tableline
\end{tabular}
\end{center}}

**Temperatures listed under the two-temperature models for these targets are not fitted but fixed at 0.98 $T(S1-S0)$ and 1.3 $T(S1-S3)$ because these targeted regions are not consistent with the general case of LTE and have $\rm{OPR} <3$. Hence these two-temperature estimates provide only a rough approximation of the temperatures and column densities present within these targets.
\tablecomments{Col.\ (1): Target.  Col.\ (2): Temperature from a one-temperature fit.  Col.\ (3): Column density
from a one-temperature fit.  Col.\ (4): Temperature of the warm component from a two-temperature fit.  Col.\ (5):
Temperature of the hot component from a two-temperature fit.  Col.\ (6): Column density of the warm component
from a two-temperature fit.  Col.\ (7): Ratio of the column density of the $J=0$ state of the hot to warm component,
$N_{0,2}/N_{0,1}$. }%Col.\ (8): Total column density of excited \htwo\ as estimated from the temperature models. }

\end{table*}

%The leading edge of N4522 (N4522\_NE)  can be fit with a temperature of 130$^{+17}_{-14}$ K for the observed distribution of 
%$S(0)$ and $S(1)$.  A two temperature fit of the same region yields a warm component with a temperature of
%134$^{+6}_{-5}$ K and a hot component of 632$^{+796}_{-495}$ K where the error intervals show the 10th and 90th bootstrap
%percentile confidence interval.  Similarly, a single temperature fit to the observed  region on the trailing side, N4522\_SW,  
%yielded a temperature of 127$^{+23}_{-18}$ K, while a two-temperature fit found  warm and hot components with temperatures 
%of 125$^{+25}_{-16}$ K and 934$^{+494}_{-808}$ K, respectively.  

%The true temperature of the warm component is likely to be between the temperatures derived from the
%single- and the two-temperature models.  

The warm \htwo\  temperature estimates listed in Table~\ref{fittemp} are compareable to those from the SINGS 
sample \citep{roussel07} as well as that
of the ram pressure-stripped spiral, ESO 137-001 \citep{sivanandam09} in the Abell 3627 cluster.
On the other hand, the temperatures of the hot components in the N4402 regions are more similar 
to the non-nuclear measurements of the nearby SINGS galaxies and are lower than that determined for ESO 137-001.
We hypothesize that the greater \htwo\ temperatures found in Abell 3627 are due to a stronger ICM pressure
being experience by ESO 137-001, relative to our sample of Virgo galaxies.

Previous  X-ray observations found an extensive amount of hot X-ray emitting gas in the tail of ESO 137-001---a good
indication of strong ICM thermal pressure  \citep{sun07,sun10}.  Other recent studies
also concluded that the general lack of X-ray tails in ram pressure-stripped Virgo cluster galaxies is indicative of the
low ICM pressure within the Virgo Cluster \citep{tonnesen11,jachym13}.  In IC 3418 (a ram pressure-stripped Virgo galaxy with the most
similar star-forming tail to that in ESO 137-001), \citet{jachym13} found the X-ray (0.5--2 keV) luminosity limit
 to be 280 times weaker than that measured in ESO 137-001 \citep{sun10}.  Shock heating was found to be the most 
likely cause of the observed enhancement in the ratio of rotational \htwo\ to IR luminosity in the 
ESO 137-001 tail \citep{sivanandam09}.  In the case where \htwo\ is heated by ram pressure-driven shocks, the
stronger ram pressure experienced by ESO 137-001 would suggest a greater amount of \htwo\ heating than in 
Virgo cluster galaxies.  The  low \htwo\ temperature that we estimate for the N4522\_SW2 region is 
also consistent with the idea that the possible shock-heating of \htwo\ in NGC 4522 is  much lower than 
that in ESO 137-001.

%Currently, we think that there may be several reasons why the leading edges of N4330, N4402 and N4501 do not 
%appear to comply with the LTE criteria for OPR=3.  %In addition, the higher energy rotational lines of our pointings have marginal 
%detections with low signal-to-noise due to severe contamination and confusion with the strong PAH emission 
%complexes.  This is also the reason why we have not modelled the excitation temperatures of these IRS targets. 
%A non-LTE condition with low OPR values indicate that there are multiple significant temperature components within the
%measured region.
% \citet{roussel07} interpret values of OPR lower than three to be a result of fluorescent excitation (in star-forming 
%regions) occuring in sufficiently low density regions where the ortho-para equilibration via collisions is incomplete.
%It is also possible that the \htwo\ molecules which formed at  T$>$200K have since cooled to T$<$200K. However, the 
%gas at the cooler stage may not have reached LTE due to insufficient density and this then results in the OPR value being
% frozen to when the gas was last in LTE at a warmer temperature \citep{sivanandam09}.  It is interesting to 
%note that the \htwo\ surface brightnesses (estimated from the sum of S(0), S(1) and S(2)) of our
%three non-LTE-compliant targets  have similarly low \htwo\ surface brightnesses to the three SINGS
%regions which also have the  lowest OPRs \citep[and hence furthest from equilibria between the
% ortho-to-para states; ][]{roussel07}.

\section{Conclusions}

In this paper, we describe the MIR spectral mapping of four Virgo spiral galaxies which show clear evidence
for strong ongoing ram pressure stripping using the IRS instrument on board  the {\it{Spitzer}} Space Telescope.
   The 8 regions that we study in this paper include the leading and trailing sides of 
the galaxies, a region in the star-forming disk as well as an extraplanar star-forming region located on the 
trailing side of NGC 4522.

A spatial coincidence between the  warm \htwo\ emission and the 8 $\mu$m PAH emission is observed for the entire
sample studied in this paper. This suggests that most of our observed warm \htwo\ is related to star formation.  
In the 
extraplanar star-forming region on the trailing side of NGC 4522 (N4522\_SW2),  we find the highest ratio of 
$S(0)$ to $S(1)$ emission, relative to the other regions in our sample.  The $S(0)/S(1)$ fraction measured in
 N4522\_SW2 is 2.6 times greater than the average fraction in all the other targeted regions positioned closer to the 
star-forming disks.  The temperature of the warm \htwo\ in N4522\_SW2 is also approximately 20\% colder than  
that of all the other regions sampled. This is probably due to the lower density of heating sources
in the extraplanar star-forming region relative to the star-forming regions within the disk of the galaxy.

%than the S(1) emission.  This S(0) emission coincides with the extraplanar star-forming regions
%of NGC 4522 which resulted from stripped gas caused by the ram pressure interaction.  It is likely that the \htwo\ gas 
%is more likely to be heated to higher order rotational energy states closer to the disk plane than further out.  
We find elevated fractions of H$_2(S0+S1+S2)$/PAH in this Virgo sample to be on average 2.6 times greater 
than that of  nearby field galaxies from the SINGS sample. This  suggests that additional mechanisms 
are contributing to  the heating of the \htwo\ molecules within our observed sample in addition to heating 
by the newly-formed stellar population.    Following the arguments for shock-heating 
in \citet{roussel07} and \citet{timmermann98}, we propose that our observed \htwo/PAH excess may be due to shocks 
generated via the ram pressure interaction.  A comparison with global PAH/TIR ratios shows that the elevated 
H$_2(S0+S1+S2)$/PAH ratios are due to increased \htwo\ emission, and not diminished PAH emission.  
Elevated fractions of \Fe/\Ne\ relative to \Si/\Sulphur\ also support the hypothesis for additional 
ISM-heating via shocks.  It is plausible that shocks are returning greater amounts of Fe from the dust grains
 back to the gas phase than Si.

Although the estimated temperatures for the warm and hot components for NGC 4402 is similar to those determined
for star-forming regions in SINGS sample \citep{roussel07}, the temperature constraints for the hot components of
N4402\_1 and N4402\_2 are much colder than that found for the ram pressure-stripped spiral, ESO 137-001, 
in the Abell 3627 cluster \citep{sivanandam09}. In contrast to the X-ray tail(s) observed in ESO 137-001, the general 
lack of of X-ray tails in Virgo galaxies experiencing ram pressure stripping suggests that the ICM pressure
is much greater in Abell 3626 than that in the Virgo Cluster \citep{sun07,sun10,tonnesen11,jachym13}. Hence,
 even though we find some evidence for shock-heating within our Virgo sample (possibly as a 
result of ram pressure),  the shock-heating is weaker in the Virgo Cluster than in Abell 3627 as a consequence
of the weaker ICM pressure.

% This is probably due to the presence of greater
%ICM pressure producing stronger shock-heating in the Abell 3627 cluster than in the Virgo Cluster. This hypothesis
%is consistent with the lack of X-ray tails in Virgo spirals experiencing ram pressure stripping as compared to 
%ESO 137-001 \citep{sun07,sun10,tonnesen11,jachym13}.

%\end{itemize}

%n conclusion, we have found an excess of warm \htwo/PAH in our sample of ram pressure stripped Virgo galaxies which is 
%approximately three times greater than that found in other nearby star-forming field galaxies.  In addition to the
%heating of \htwo\ via the young stellar population, we also attribute part of the observed \htwo/PAH excess to the 
%possible shocks and turbulences which may have been generated through the ram pressure interaction.

\acknowledgments

This work is based in part on observations made with the {\it{Spitzer}} Space
Telescope, which
is
operated by the Jet Propulsion Laboratory, California Institute of Technology
under a contract with NASA. Support for this work was provided by NASA through
a grant (R09078) issued by JPL/Caltech.  OIW is the recipient of a Super Science Fellowship from the Australian
Research Council. The authors also thank the anonymous referee
for the constructive comments which have greatly improved this paper.

\bibliographystyle{apj}
\bibliography{mn-jour,paperef}

\begin{thebibliography}{73}
\expandafter\ifx\csname natexlab\endcsname\relax\def\natexlab#1{#1}\fi

\bibitem[{{Abramson} {et~al.}(2011){Abramson}, {Kenney}, {Crowl}, {Chung}, {van
  Gorkom}, {Vollmer}, \& {Schiminovich}}]{abramson11}
{Abramson}, A., {Kenney}, J.~D.~P., {Crowl}, H.~H., {Chung}, A., {van Gorkom},
  J.~H., {Vollmer}, B., \& {Schiminovich}, D. 2011, AJ, 141, 164

\bibitem[{{Alonso-Herrero} {et~al.}(2003){Alonso-Herrero}, {Rieke}, {Rieke}, \&
  {Kelly}}]{alonso03}
{Alonso-Herrero}, A., {Rieke}, G.~H., {Rieke}, M.~J., \& {Kelly}, D.~M. 2003,
  AJ, 125, 1210

\bibitem[{{Alonso-Herrero} {et~al.}(1997){Alonso-Herrero}, {Rieke}, {Rieke}, \&
  {Ruiz}}]{alonso97}
{Alonso-Herrero}, A., {Rieke}, M.~J., {Rieke}, G.~H., \& {Ruiz}, M. 1997, \apj,
  482, 747

\bibitem[{{Appleton} {et~al.}(2006){Appleton}, {Xu}, {Reach}, {Dopita}, {Gao},
  {Lu}, {Popescu}, {Sulentic}, {Tuffs}, \& {Yun}}]{appleton06}
{Appleton}, P.~N., {et~al.} 2006, ApJ, 639, L51

\bibitem[{{Armus} {et~al.}(2007){Armus}, {Charmandaris}, {Bernard-Salas},
  {Spoon}, {Marshall}, {Higdon}, {Desai}, {Teplitz}, {Hao}, {Devost}, {Brandl},
  {Wu}, {Sloan}, {Soifer}, {Houck}, \& {Herter}}]{armus07}
{Armus}, L., {et~al.} 2007, ApJ, 656, 148

\bibitem[{{Bendo} {et~al.}(2008){Bendo}, {Draine}, {Engelbracht}, {Helou},
  {Thornley}, {Bot}, {Buckalew}, {Calzetti}, {Dale}, {Hollenbach}, {Li}, \&
  {Moustakas}}]{bendo08}
{Bendo}, G.~J., {et~al.} 2008, MNRAS, 389, 629

\bibitem[{{Burton} {et~al.}(1992){Burton}, {Hollenbach}, \&
  {Tielens}}]{burton92}
{Burton}, M.~G., {Hollenbach}, D.~J., \& {Tielens}, A.~G.~G. 1992, ApJ, 399,
  563

\bibitem[{{Chung} {et~al.}(2009){Chung}, {van Gorkom}, {Kenney}, {Crowl}, \&
  {Vollmer}}]{chung09}
{Chung}, A., {van Gorkom}, J.~H., {Kenney}, J.~D.~P., {Crowl}, H., \&
  {Vollmer}, B. 2009, AJ, 138, 1741

\bibitem[{{Chung} {et~al.}(2007){Chung}, {van Gorkom}, {Kenney}, \&
  {Vollmer}}]{chung07}
{Chung}, A., {van Gorkom}, J.~H., {Kenney}, J.~D.~P., \& {Vollmer}, B. 2007,
  ApJ, 659, L115

\bibitem[{{Cortese} {et~al.}(2010){Cortese}, {Davies}, {Pohlen}, {Baes},
  {Bendo}, {Bianchi}, {Boselli}, {de Looze}, {Fritz}, {Verstappen}, {Bomans},
  {Clemens}, {Corbelli}, {Dariush}, {di Serego Alighieri}, {Fadda},
  {Garcia-Appadoo}, {Gavazzi}, {Giovanardi}, {Grossi}, {Hughes}, {Hunt},
  {Jones}, {Madden}, {Pierini}, {Sabatini}, {Smith}, {Vlahakis}, {Xilouris}, \&
  {Zibetti}}]{cortese10}
{Cortese}, L., {et~al.} 2010, A\&A, 518, L49

\bibitem[{{Cortese} {et~al.}(2007){Cortese}, {Marcillac}, {Richard},
  {Bravo-Alfaro}, {Kneib}, {Rieke}, {Covone}, {Egami}, {Rigby}, {Czoske}, \&
  {Davies}}]{cortese07}
---. 2007, MNRAS, 376, 157

\bibitem[{{Crowl} \& {Kenney}(2006)}]{crowl06}
{Crowl}, H.~H., \& {Kenney}, J.~D.~P. 2006, ApJ, 649, L75

\bibitem[{{Crowl} \& {Kenney}(2008)}]{crowl08}
---. 2008, AJ, 136, 1623

\bibitem[{{Crowl} {et~al.}(2010){Crowl}, {Kenney}, {Chung}, {Blanton}, \& {van
  Gorkom}}]{crowl10}
{Crowl}, H.~H., {Kenney}, J.~D.~P., {Chung}, A., {Blanton}, M., \& {van
  Gorkom}, J.~H. 2010, in Astronomical Society of the Pacific Conference
  Series, Vol. 423, Galaxy Wars: Stellar Populations and Star Formation in
  Interacting Galaxies, ed. B.~{Smith}, J.~{Higdon}, S.~{Higdon}, \&
  N.~{Bastian}, 97

\bibitem[{{Crowl} {et~al.}(2005){Crowl}, {Kenney}, {van Gorkom}, \&
  {Vollmer}}]{crowl05}
{Crowl}, H.~H., {Kenney}, J.~D.~P., {van Gorkom}, J.~H., \& {Vollmer}, B. 2005,
  AJ, 130, 65

\bibitem[{{Dale} {et~al.}(2007){Dale}, {Gil de Paz}, {Gordon}, {Hanson},
  {Armus}, {Bendo}, {Bianchi}, {Block}, {Boissier}, {Boselli}, {Buckalew},
  {Buat}, {Burgarella}, {Calzetti}, {Cannon}, {Engelbracht}, {Helou},
  {Hollenbach}, {Jarrett}, {Kennicutt}, {Leitherer}, {Li}, {Madore}, {Martin},
  {Meyer}, {Murphy}, {Regan}, {Roussel}, {Smith}, {Sosey}, {Thilker}, \&
  {Walter}}]{dale07}
{Dale}, D.~A., {et~al.} 2007, ApJ, 655, 863

\bibitem[{{Dale} {et~al.}(2009){Dale}, {Smith}, {Schlawin}, {Armus},
  {Buckalew}, {Cohen}, {Helou}, {Jarrett}, {Johnson}, {Moustakas}, {Murphy},
  {Roussel}, {Sheth}, {Staudaher}, {Bot}, {Calzetti}, {Engelbracht}, {Gordon},
  {Hollenbach}, {Kennicutt}, \& {Malhotra}}]{dale09}
---. 2009, ApJ, 693, 1821

\bibitem[{{de Jong} {et~al.}(1985){de Jong}, {Klein}, {Wielebinski}, \&
  {Wunderlich}}]{dejong85}
{de Jong}, T., {Klein}, U., {Wielebinski}, R., \& {Wunderlich}, E. 1985, A\&A,
  147, L6

\bibitem[{{de Vaucouleurs} {et~al.}(1991){de Vaucouleurs}, {de Vaucouleurs},
  {Corwin}, {Buta}, {Paturel}, \& {Fouque}}]{devaucouleurs91}
{de Vaucouleurs}, G., {de Vaucouleurs}, A., {Corwin}, Jr., H.~G., {Buta},
  R.~J., {Paturel}, G., \& {Fouque}, P. 1991, {Third Reference Catalogue of
  Bright Galaxies} (Springer-Verlag Berlin Heidelberg New York)

\bibitem[{{Desert} \& {Dennefeld}(1988)}]{desert88}
{Desert}, F.~X., \& {Dennefeld}, M. 1988, A\&A, 206, 227

\bibitem[{{Draine}(2004)}]{draine04}
{Draine}, B.~T. 2004, {Astrophysics of Dust in Cold Clouds}, 213

\bibitem[{{Dumas} {et~al.}(2011){Dumas}, {Schinnerer}, {Tabatabaei}, {Beck},
  {Velusamy}, \& {Murphy}}]{dumas11}
{Dumas}, G., {Schinnerer}, E., {Tabatabaei}, F.~S., {Beck}, R., {Velusamy}, T.,
  \& {Murphy}, E. 2011, AJ, 141, 41

\bibitem[{{Falgarone} {et~al.}(2005){Falgarone}, {Verstraete}, {Pineau Des
  For{\^e}ts}, \& {Hily-Blant}}]{falgarone05}
{Falgarone}, E., {Verstraete}, L., {Pineau Des For{\^e}ts}, G., \&
  {Hily-Blant}, P. 2005, A\&A, 433, 997

\bibitem[{{Forbes} \& {Ward}(1993)}]{forbes93}
{Forbes}, D.~A., \& {Ward}, M.~J. 1993, ApJ, 416, 150

\bibitem[{{Gavazzi} \& {Jaffe}(1987)}]{gavazzi87}
{Gavazzi}, G., \& {Jaffe}, W. 1987, A\&A, 186, L1

\bibitem[{{Giard} {et~al.}(1994){Giard}, {Bernard}, {Lacombe}, {Normand}, \&
  {Rouan}}]{giard94}
{Giard}, M., {Bernard}, J.~P., {Lacombe}, F., {Normand}, P., \& {Rouan}, D.
  1994, A\&A, 291, 239

\bibitem[{{Haas} {et~al.}(2002){Haas}, {Klaas}, \& {Bianchi}}]{haas02}
{Haas}, M., {Klaas}, U., \& {Bianchi}, S. 2002, A\&A, 385, L23

\bibitem[{{Helou} {et~al.}(2001){Helou}, {Malhotra}, {Hollenbach}, {Dale}, \&
  {Contursi}}]{helou01}
{Helou}, G., {Malhotra}, S., {Hollenbach}, D.~J., {Dale}, D.~A., \& {Contursi},
  A. 2001, ApJ, 548, L73

\bibitem[{{Helou} {et~al.}(1985){Helou}, {Soifer}, \&
  {Rowan-Robinson}}]{helou85}
{Helou}, G., {Soifer}, B.~T., \& {Rowan-Robinson}, M. 1985, ApJ, 298, L7

\bibitem[{{Helou}(2004)}]{helou04}
{Helou}, G.~a. 2004, ApJS, 154, 253

\bibitem[{{Higdon} {et~al.}(2006){Higdon}, {Armus}, {Higdon}, {Soifer}, \&
  {Spoon}}]{higdon06}
{Higdon}, S.~J.~U., {Armus}, L., {Higdon}, J.~L., {Soifer}, B.~T., \& {Spoon},
  H.~W.~W. 2006, ApJ, 648, 323

\bibitem[{{Hoernes} {et~al.}(1998){Hoernes}, {Berkhuijsen}, \&
  {Xu}}]{hoernes98}
{Hoernes}, P., {Berkhuijsen}, E.~M., \& {Xu}, C. 1998, A\&A, 334, 57

\bibitem[{{Houck} {et~al.}(2004){Houck}, {Roellig}, {van Cleve}, {Forrest},
  {Herter}, {Lawrence}, {Matthews}, {Reitsema}, {Soifer}, {Watson}, {Weedman},
  {Huisjen}, {Troeltzsch}, {Barry}, {Bernard-Salas}, {Blacken}, {Brandl},
  {Charmandaris}, {Devost}, {Gull}, {Hall}, {Henderson}, {Higdon}, {Pirger},
  {Schoenwald}, {Sloan}, {Uchida}, {Appleton}, {Armus}, {Burgdorf},
  {Fajardo-Acosta}, {Grillmair}, {Ingalls}, {Morris}, \& {Teplitz}}]{houck04}
{Houck}, J.~R., {et~al.} 2004, ApJS, 154, 18

\bibitem[{{Hughes} {et~al.}(2006){Hughes}, {Wong}, {Ekers}, {Staveley-Smith},
  {Filipovic}, {Maddison}, {Fukui}, \& {Mizuno}}]{hughes06}
{Hughes}, A., {Wong}, T., {Ekers}, R., {Staveley-Smith}, L., {Filipovic}, M.,
  {Maddison}, S., {Fukui}, Y., \& {Mizuno}, N. 2006, MNRAS, 370, 363

\bibitem[{{J{\'a}chym} {et~al.}(2013){J{\'a}chym}, {Kenney}, {R{\v z}ui{\v
  c}ka}, {Sun}, {Combes}, \& {Palou{\v s}}}]{jachym13}
{J{\'a}chym}, P., {Kenney}, J.~D.~P., {R{\v z}ui{\v c}ka}, A., {Sun}, M.,
  {Combes}, F., \& {Palou{\v s}}, J. 2013, A\&A, 556, A99

\bibitem[{{Kaufman} {et~al.}(2006){Kaufman}, {Wolfire}, \&
  {Hollenbach}}]{kaufman06}
{Kaufman}, M.~J., {Wolfire}, M.~G., \& {Hollenbach}, D.~J. 2006, ApJ, 644, 283

\bibitem[{{Kenney} {et~al.}(2014){Kenney}, {Geha}, {J{\'a}chym}, {Crowl},
  {Dague}, {Chung}, {van Gorkom}, \& {Vollmer}}]{kenney14}
{Kenney}, J.~D.~P., {Geha}, M., {J{\'a}chym}, P., {Crowl}, H.~H., {Dague}, W.,
  {Chung}, A., {van Gorkom}, J., \& {Vollmer}, B. 2014, ApJ, 780, 119

\bibitem[{{Kenney} {et~al.}(2004){Kenney}, {van Gorkom}, \&
  {Vollmer}}]{kenney04}
{Kenney}, J.~D.~P., {van Gorkom}, J.~H., \& {Vollmer}, B. 2004, AJ, 127, 3361

\bibitem[{{Kennicutt} {et~al.}(2003){Kennicutt}, {Armus}, {Bendo}, {Calzetti},
  {Dale}, {Draine}, {Engelbracht}, {Gordon}, {Grauer}, {Helou}, {Hollenbach},
  {Jarrett}, {Kewley}, {Leitherer}, {Li}, {Malhotra}, {Regan}, {Rieke},
  {Rieke}, {Roussel}, {Smith}, {Thornley}, \& {Walter}}]{kennicutt03}
{Kennicutt}, Jr., R.~C., {et~al.} 2003, PASP, 115, 928

\bibitem[{{Koopmann} \& {Kenney}(2004)}]{koopmann04}
{Koopmann}, R.~A., \& {Kenney}, J.~D.~P. 2004, ApJ, 613, 866

\bibitem[{{Koopmann} {et~al.}(2001){Koopmann}, {Kenney}, \&
  {Young}}]{koopmann01}
{Koopmann}, R.~A., {Kenney}, J.~D.~P., \& {Young}, J. 2001, ApJS, 135, 125

\bibitem[{{Kormendy} \& {Bender}(2012)}]{kormendy12}
{Kormendy}, J., \& {Bender}, R. 2012, ApJS, 198, 2

\bibitem[{{Merluzzi} {et~al.}(2013){Merluzzi}, {Busarello}, {Dopita}, {Haines},
  {Steinhauser}, {Mercurio}, {Rifatto}, {Smith}, \& {Schindler}}]{merluzzi13}
{Merluzzi}, P., {et~al.} 2013, MNRAS, 429, 1747

\bibitem[{{Miller} \& {Owen}(2001)}]{miller01}
{Miller}, N.~A., \& {Owen}, F.~N. 2001, AJ, 121, 1903

\bibitem[{{Murphy} {et~al.}(2006){Murphy}, {Helou}, {Braun}, {Kenney}, {Armus},
  {Calzetti}, {Draine}, {Kennicutt}, {Roussel}, {Walter}, {Bendo}, {Buckalew},
  {Dale}, {Engelbracht}, {Smith}, \& {Thornley}}]{murphy06}
{Murphy}, E.~J., {et~al.} 2006, ApJ, 651, L111

\bibitem[{{Murphy} {et~al.}(2008){Murphy}, {Helou}, {Kenney}, {Armus}, \&
  {Braun}}]{murphy08}
{Murphy}, E.~J., {Helou}, G., {Kenney}, J.~D.~P., {Armus}, L., \& {Braun}, R.
  2008, \apj, 678, 828

\bibitem[{{Murphy} {et~al.}(2009){Murphy}, {Kenney}, {Helou}, {Chung}, \&
  {Howell}}]{murphy09}
{Murphy}, E.~J., {Kenney}, J.~D.~P., {Helou}, G., {Chung}, A., \& {Howell},
  J.~H. 2009, ApJ, 694, 1435

\bibitem[{{O'Halloran} {et~al.}(2006){O'Halloran}, {Satyapal}, \&
  {Dudik}}]{ohalloran06}
{O'Halloran}, B., {Satyapal}, S., \& {Dudik}, R.~P. 2006, ApJ, 641, 795

\bibitem[{{Pfrommer} \& {Dursi}(2010)}]{pfrommer10}
{Pfrommer}, C., \& {Dursi}, J.~L. 2010, Nat, 6, 520

\bibitem[{{Reddy} \& {Yun}(2004)}]{reddy04}
{Reddy}, N.~A., \& {Yun}, M.~S. 2004, ApJ, 600, 695

\bibitem[{{Rigopoulou} {et~al.}(2002){Rigopoulou}, {Kunze}, {Lutz}, {Genzel},
  \& {Moorwood}}]{rigopoulou02}
{Rigopoulou}, D., {Kunze}, D., {Lutz}, D., {Genzel}, R., \& {Moorwood},
  A.~F.~M. 2002, A\&A, 389, 374

\bibitem[{{Roussel} {et~al.}(2007){Roussel}, {Helou}, {Hollenbach}, {Draine},
  {Smith}, {Armus}, {Schinnerer}, {Walter}, {Engelbracht}, {Thornley},
  {Kennicutt}, {Calzetti}, {Dale}, {Murphy}, \& {Bot}}]{roussel07}
{Roussel}, H., {et~al.} 2007, ApJ, 669, 959

\bibitem[{{Savage} \& {Sembach}(1996)}]{savage96}
{Savage}, B.~D., \& {Sembach}, K.~R. 1996, ApJ, 470, 893

\bibitem[{{Sivanandam} {et~al.}(2010){Sivanandam}, {Rieke}, \&
  {Rieke}}]{sivanandam09}
{Sivanandam}, S., {Rieke}, M.~J., \& {Rieke}, G.~H. 2010, \apj, 717, 147

\bibitem[{{Smith} {et~al.}(2007{\natexlab{a}}){Smith}, {Armus}, {Dale},
  {Roussel}, {Sheth}, {Buckalew}, {Jarrett}, {Helou}, \&
  {Kennicutt}}]{smith07a}
{Smith}, J.~D.~T., {et~al.} 2007{\natexlab{a}}, PASP, 119, 1133

\bibitem[{{Smith} {et~al.}(2007{\natexlab{b}}){Smith}, {Draine}, {Dale},
  {Moustakas}, {Kennicutt}, {Helou}, {Armus}, {Roussel}, {Sheth}, {Bendo},
  {Buckalew}, {Calzetti}, {Engelbracht}, {Gordon}, {Hollenbach}, {Li},
  {Malhotra}, {Murphy}, \& {Walter}}]{smith07b}
---. 2007{\natexlab{b}}, ApJ, 656, 770

\bibitem[{{Smith} {et~al.}(2010){Smith}, {Lucey}, {Hammer}, {Hornschemeier},
  {Carter}, {Hudson}, {Marzke}, {Mouhcine}, {Eftekharzadeh}, {James},
  {Khosroshahi}, {Kourkchi}, \& {Karick}}]{smith10}
{Smith}, R.~J., {et~al.} 2010, MNRAS, 408, 1417

\bibitem[{{Sun} {et~al.}(2010){Sun}, {Donahue}, {Roediger}, {Nulsen}, {Voit},
  {Sarazin}, {Forman}, \& {Jones}}]{sun10}
{Sun}, M., {Donahue}, M., {Roediger}, E., {Nulsen}, P.~E.~J., {Voit}, G.~M.,
  {Sarazin}, C., {Forman}, W., \& {Jones}, C. 2010, ApJ, 708, 946

\bibitem[{{Sun} {et~al.}(2007){Sun}, {Donahue}, \& {Voit}}]{sun07}
{Sun}, M., {Donahue}, M., \& {Voit}, G.~M. 2007, ApJ, 671, 190

\bibitem[{{Sun} {et~al.}(2006){Sun}, {Jones}, {Forman}, {Nulsen}, {Donahue}, \&
  {Voit}}]{sun06}
{Sun}, M., {Jones}, C., {Forman}, W., {Nulsen}, P.~E.~J., {Donahue}, M., \&
  {Voit}, G.~M. 2006, ApJ, 637, L81

\bibitem[{{Tielens} {et~al.}(1999){Tielens}, {Hony}, {van Kerckhoven}, \&
  {Peeters}}]{tielens99}
{Tielens}, A.~G.~G.~M., {Hony}, S., {van Kerckhoven}, C., \& {Peeters}, E.
  1999, in ESA Special Publication, Vol. 427, The Universe as Seen by ISO, ed.
  P.~{Cox} \& M.~{Kessler}, 579

\bibitem[{{Timmermann}(1998)}]{timmermann98}
{Timmermann}, R. 1998, ApJ, 498, 246

\bibitem[{{Tonnesen} \& {Bryan}(2012)}]{tonnesen12}
{Tonnesen}, S., \& {Bryan}, G.~L. 2012, MNRAS, 422, 1609

\bibitem[{{Tonnesen} {et~al.}(2011){Tonnesen}, {Bryan}, \& {Chen}}]{tonnesen11}
{Tonnesen}, S., {Bryan}, G.~L., \& {Chen}, R. 2011, ApJ, 731, 98

\bibitem[{{Voit}(1992)}]{voit92}
{Voit}, G.~M. 1992, MNRAS, 258, 841

\bibitem[{{Vollmer} {et~al.}(2004){Vollmer}, {Beck}, {Kenney}, \& {van
  Gorkom}}]{vollmer04}
{Vollmer}, B., {Beck}, R., {Kenney}, J.~D.~P., \& {van Gorkom}, J.~H. 2004, AJ,
  127, 3375

\bibitem[{{Vollmer} \& {Leroy}(2011)}]{vollmer11}
{Vollmer}, B., \& {Leroy}, A.~K. 2011, AJ, 141, 24

\bibitem[{{Vollmer} {et~al.}(2007){Vollmer}, {Soida}, {Beck}, {Urbanik},
  {Chy{\.z}y}, {Otmianowska-Mazur}, {Kenney}, \& {van Gorkom}}]{vollmer07}
{Vollmer}, B., {Soida}, M., {Beck}, R., {Urbanik}, M., {Chy{\.z}y}, K.~T.,
  {Otmianowska-Mazur}, K., {Kenney}, J.~D.~P., \& {van Gorkom}, J.~H. 2007,
  A\&A, 464, L37

\bibitem[{{Vollmer} {et~al.}(2008){Vollmer}, {Soida}, {Chung}, {van Gorkom},
  {Otmianowska-Mazur}, {Beck}, {Urbanik}, \& {Kenney}}]{vollmer08}
{Vollmer}, B., {Soida}, M., {Chung}, A., {van Gorkom}, J.~H.,
  {Otmianowska-Mazur}, K., {Beck}, R., {Urbanik}, M., \& {Kenney}, J.~D.~P.
  2008, A\&A, 483, 89

\bibitem[{{Wong} \& {Kenney}(2014)}]{wong10b}
{Wong}, O.~I., \& {Kenney}, J.~D.~P. 2014, in preparation

\bibitem[{{Wong} {et~al.}(2014){Wong}, {Kenney}, {Howell}, {Murphy}, \&
  {Helou}}]{wong10a}
{Wong}, O.~I., {Kenney}, J.~D.~P., {Howell}, J.~H., {Murphy}, E.~J., \&
  {Helou}, G. 2014, in preparation

\bibitem[{{Yagi} {et~al.}(2010){Yagi}, {Yoshida}, {Komiyama}, {Kashikawa},
  {Furusawa}, {Okamura}, {Graham}, {Miller}, {Carter}, {Mobasher}, \&
  {Jogee}}]{yagi10}
{Yagi}, M., {et~al.} 2010, AJ, 140, 1814

\bibitem[{{Yun} {et~al.}(2001){Yun}, {Reddy}, \& {Condon}}]{yun01}
{Yun}, M.~S., {Reddy}, N.~A., \& {Condon}, J.~J. 2001, ApJ, 554, 803

\end{thebibliography}

\end{document}